\documentclass[10pt,twocolumn]{IEEEtran} 

\usepackage{amsmath,xurl}

\usepackage{epsfig,amssymb,amsbsy,verbatim,array,enumerate}
\usepackage{pstricks,psfrag,theorem,cite,paralist,bm,subcaption}
\usepackage{enumitem,wasysym}

\let\intern=\iftrue

\newcommand{\argmin}{\operatornamewithlimits{arg\ min}}
\def\figref#1{Fig.\,\ref{#1}}%
\def\E{\mathbb{E}}
\def\P{\mathbb{P}}
\def\R{\mathbb{R}}

\def\N{\mathbb{N}}

\def\C{\mathbb{C}}

\def\calH{\mathcal{H}}
\def\calC{\mathcal{C}}

\def\calX{\mathcal{X}}

\def\calQ{\mathcal{Q}}
\def\calH{\mathcal{H}}

\def\ie{{\em i.e.}}
\def\eg{{\em e.g.}}

\def\var{\operatorname{var}}

\def\sir{\mathsf{SIR}}

\def\sfr{\mathsf{SF}}

\def\x{\mathsf{x}}
\def\y{\mathsf{y}}
\def\z{\mathsf{z}}
\def\w{\mathsf{w}}
\def\q{\mathsf{q}}

\def\mh#1{{#1^\oplus}}
\def\imh#1{{#1^\ominus}}

\def\sinc{\operatorname{sinc}}
\def\dd{\mathrm{d}}

\def\one{\mathbf{1}}

\def\phi{\varphi}
\def\md{\bar F_{[\sir]}}
\def\na{\breve{n}}

\newcommand{\iu}{\mathrm{i}}

\newtheorem{fact}{Fact}
\newtheorem{theorem}{Theorem}
\newtheorem{lemma}{Lemma}
\newtheorem{corollary}{Corollary}
\newtheorem{definition}{Definition}

\newtheorem{recipe}{Recipe}

\newlength{\figwidth}


\begin{document}
\title{Q Cells in Wireless Networks} 
\author{Martin Haenggi, \IEEEmembership{Fellow, IEEE}
\thanks{Martin Haenggi is with the Dept.~of Electrical Engineering, University of Notre Dame, Indiana, USA. E-mail: {\tt mhaenggi@nd.edu}.}}
\maketitle
\begin{abstract}
For a given set of transmitters such as cellular base stations or WiFi access points, is it possible
to analytically characterize the set of locations that are ``covered" 
in the sense that users at these locations experience a certain minimum quality of service?
In this paper, we affirmatively answer this question, by providing explicit simple outer bounds and
estimates for the coverage manifold. The key geometric elements of our analytical method are
the {\em Q cells}, defined as the intersections of a small number of disks. The Q cell of a transmitter
is an outer bound to the service region of the transmitter, and, in turn, the union of Q cells is
an outer bound to the coverage manifold. In infinite networks, connections to the
meta distribution of the signal-to-interference ratio allow for a scaling of the Q cells to obtain accurate
estimates of the coverage manifold.
\end{abstract}
\captionsetup{font=footnotesize}
\section{Introduction}
\subsection{Problem Formulation}
Coverage is a key performance metric of wireless networks. It has been studied extensively in terms of
the fraction of the area of a region that is ``covered" by wireless service, see, \eg, \cite{net:Baccelli09now,net:Haenggi18book} and references therein.
This paper addresses the more challenging
problem of analytically characterizing the {\em coverage manifold} $\calC\subset\R^2$, 
which is the subset of the plane that is covered in the sense that the probability that the signal-to-interference ratio
(SIR) exceeds a threshold $\theta$ is at least $u$.
Specifically, for a set of transmitter locations, the problem is to find tight bounds and estimates for $\calC$ that
are described using only elementary geometric shapes---polygons and disk segments---and can be calculated
very efficiently.

\subsection{Approach and Related Work}
\label{sec:approach}
Our approach is based on the so-called {\em Q cells}, which consists of the locations in which a serving
transmitter provides enough power such that no single interferer can cause the location to be uncovered.
Conversely, for any location outside the Q cell of a transmitter, there exists an interferer that causes the 
location to be uncovered.

Q cells generalize the standard Voronoi cells and provide an outer bound on the region that is covered by a
transmitter. They are related to the SIR cells introduced in \cite[Ch.~5]{net:Baccelli09now}, 
defined as the set of locations where the SIR exceeds a certain threshold.
The SIR cells are random sets that depend on the instantaneous values of the fading coefficients
in the propagation model, which results in a much more complex and time-dependent structure. A
location may be inside an SIR cell at a given time but outside just a fraction of a second later, and the opposite
may hold for a location merely a wavelength apart. In contrast, the Q cells are deterministic (for a given set of
transmitters) and do not depend on time. 
As such, they correspond to the regions that are shown in coverage maps.

Simplified interference modeling using only the nearest interferer(s) dates back to at least \cite{net:Gilhousen91tvt}
and \cite{net:Wyner94}, whose model became known as the ``Wyner model".
More recently,
\cite{net:Weber10tcom} used the concept of {\em dominant nodes} to
bound the so-called transmission capacity of ad hoc
networks, where an interferer is said to be dominant if its interference alone can cause an outage.
In this terminology, our Q cells are defined as the locations for which there are no dominant interferers.
In \cite{net:Kusula13wcl}, the total interference is approximated using the nearest interferer only, and
\cite{net:Shafie21jsac} uses a dominant-interferer approach for THz networks, with the additional
assumption that non-dominant interferers by themselves cannot causes outages.
Considering only the dominant interferer results can yield asymptotically exact or scaling results in terms of
the tail of the interference or in the low-outage regime, as established in \cite{net:Mordachev09jsac,net:Lee18tit}.

Another line of work models the additional interference from non-dominant interferers by
adding a deterministic term \cite{net:Qin23tcom} or a Gaussian random variable \cite{net:Chetlur17tcom}.
This is essential for path loss exponents near $2$, where the interference from distant transmitters is
significant \cite{net:stogblog-simalysis}.
These articles
focus on a form of coverage probability or coverage area fraction, not estimates of the coverage manifold.

Related in their objective but rather different in their approach
are the works on using machine learning techniques for coverage estimation,
using convolutional neural networks \cite{net:Mondal22tccn} and generative adversarial networks \cite{net:Mondal24nl}.
They require large data sets and extensive training, cannot be use to obtain bounds,
and do not lend themselves for simple interpretations,
in contrast to our direct geometric method.

The SIR meta distribution (MD) \cite{net:Haenggi16twc,net:Haenggi21cl1} is also relevant for our work since
it equals the fraction of the network area that is covered in infinite networks where the transmitters are modeled as a point process. We will take advantage of this connection to turn the coverage manifold bound that is
based on Q cells into an accurate estimate of $\calC$. Conversely, if the MD is not available,
it can be upper bounded by the area fraction covered by the Q cells.
Of particular interest is the notion of {\em separability} \cite{net:Feng20twc} for Poisson networks,
which asserts that the MD separates into the product of a function of $\theta$ and a function of $u$ for
a certain set of values of $(\theta,u)$. The separable regime corresponds to
the {\em stringent QoS regime} in this paper, which is the regime of practical interest for the quality-of-service (QoS)
parameters $\theta$ and $u$.

\subsection{Contributions}
\begin{itemize}
\item We show that the QoS parameters $\theta$ and $u$ can be lumped together into a single {\em stringency} parameter $\sigma$. Based on $\sigma$, we define lax and stringent QoS regimes.
\item We introduce {\em Q cells}, which depend on $\sigma$ and generalize Voronoi cells.
They are defined as the intersection of disks given by
of M\"obius transformed half-planes (instead of the intersection of half-planes forming the Voronoi cells), and
their union outer bounds the coverage manifold (Thm.~\ref{thm:main}).
\item Focusing on locations with two equidistant interferers, we show that cutting off corner pieces
of the Q cell results in {\em refined Q cells} that are tighter outer bounds on the coverage regions
(Thm.~\ref{thm:main2}).
\item For infinite networks, we give a simple universal lower bound on the fraction of uncovered users 
(Thm.~\ref{thm:uncov}) and
explore and exploit the connection to the SIR MD, which leads to a universal
upper bound on the MD (Cor.~\ref{cor:md_bound}) and permits the scaling of Q cells to
obtain accurate estimates of the coverage manifold (Recipe \ref{recipe}).
\end{itemize}

\section{Preliminaries}
\subsection{Notation}
Since the map $x\mapsto x/(1+x)$, introduced in \cite{net:Haenggi20cl} as the MH (M\"obius homeomorphic) transform, and its inverse appear frequently throughout the paper, we use the compact notation 
\begin{align*}
   \mh{x}&\triangleq x/(1+x),\; x\in\R\setminus\{-1\}\\
    \imh{x}&\triangleq x/(1-x),\; x\in\R\setminus\{1\} \,.
   \label{mh}
\end{align*}
Important identities include $\imh{(\mh{x})}\equiv x$ and $\mh{(1/x)}\equiv 1-\mh{x}$.

The open disk centered at $x$ with radius $r$ is denoted by $b(x,r)$, the cumulative distribution function (cdf)
of a random variable $X$ is $F_X$, its complement (ccdf) is $\bar F_X$, and its inverse (quantile function)
is $F^{-1}_H$. For $n\in\N$, $[n]\triangleq \{1,2,\ldots,n\}$.
For an uncountable set $B\subset\R^d$, $\partial B$ is its boundary, and $|B|$ is its
$d$-dimensional Lebesgue measure.
For a countable set $\calX$, $|\calX|$ is its cardinality.

For the standard mapping from $\R^2\mapsto \C$ and its inverse, we use ${}^\sharp$ and ${}^\flat$, respectively, \ie, for $(x,y)\in\R^2$,
$(x,y)^\sharp\triangleq x+\iu y$ where $\iu^2=-1$, and for $z\in\C$, $z^\flat\triangleq(\Re(z),\Im(z))$. The origin in $\R^2$ is denoted as $o=0^\flat$. 

\subsection{Three Circular Lemmas}
\begin{lemma}[Intersections of disks]
The intersection of $n$ disks 
\[ Q=\bigcap_{i\in[n]} b(x_i,r_i) \]
is either empty or has a boundary $\partial Q$ formed by $1\leq \na\leq 2(n-1)$ arcs.
If $\breve{n}=1$, it is a single disk,
if $\breve{n}=2$, $Q$ consists of two disk segments, and
if $\breve{n}>2$, it consists of a convex polygon plus $\na$ disk segments.
\label{lem:inter}
\end{lemma}
\begin{IEEEproof}
Follows from elementary geometry.
\end{IEEEproof}

\begin{lemma}[Equal distance ratio circle]
The set of locations $(x,y)\in\R^2$ such that the distance ratio to the two
points $(0,0)$ and $(b,0)$ equals $\rho\neq 1$
describes a circle with center $(c(\rho),0)$ where $c(\rho)=b/(1-\rho^2)$ and radius $r(\rho)=|(b/\rho)\imh{(\rho^2)}|=\rho |c(\rho)|$.
For $\rho=1$ (equidistance), the locations form the line $x=b/2$.
\label{lem:edrc}
\end{lemma}
\begin{IEEEproof}
The prescribed distance ratio implies
\[ \frac{(x-b)^2+y^2}{x^2+y^2} = \rho^2, \]
which, for $\rho\neq 1$, can be rearranged to 
\[ \left(x-\frac b{1-\rho^2}\right)^2+y^2=\frac{b^2\rho^2}{(1-\rho^2)^2} .\]
This is the equation for a circle with the given radius and center.
Re-writing the expression for the center as $c(\rho)=(b/\rho^2)\imh{(\rho^2)}$
reveals that center and radius are related by a factor $\pm\rho$.
For $\rho=1$, the equidistant locations form the line $x=b/2$.
\end{IEEEproof}
{\em Remarks.}
\begin{itemize}
\item If $\rho$ is replaced by $1/\rho$, the radius stays the same, and the center point is reflected
at $x=b/2$, \ie, $c(1/\rho)=b-c(\rho)$.
\item The result is easily extended  to general locations $\x$ and $\x'$ instead of $(0,0)$ and $(b,0)$,
respectively, by setting $b=\|\x'-\x\|$, and placing the circle center at $\x+\frac{c(\rho)}b (\x'-\x)$.

\item The insight that the set of locations
at which the distance ratio to two fixed points is constant is a circle dates back more than 2200
years---the Greek mathematician Apollonius defined a circle in this way \cite[Ch.~2]{net:Ogilvy69book}.
\end{itemize}

\begin{figure}
\centerline{\epsfig{file=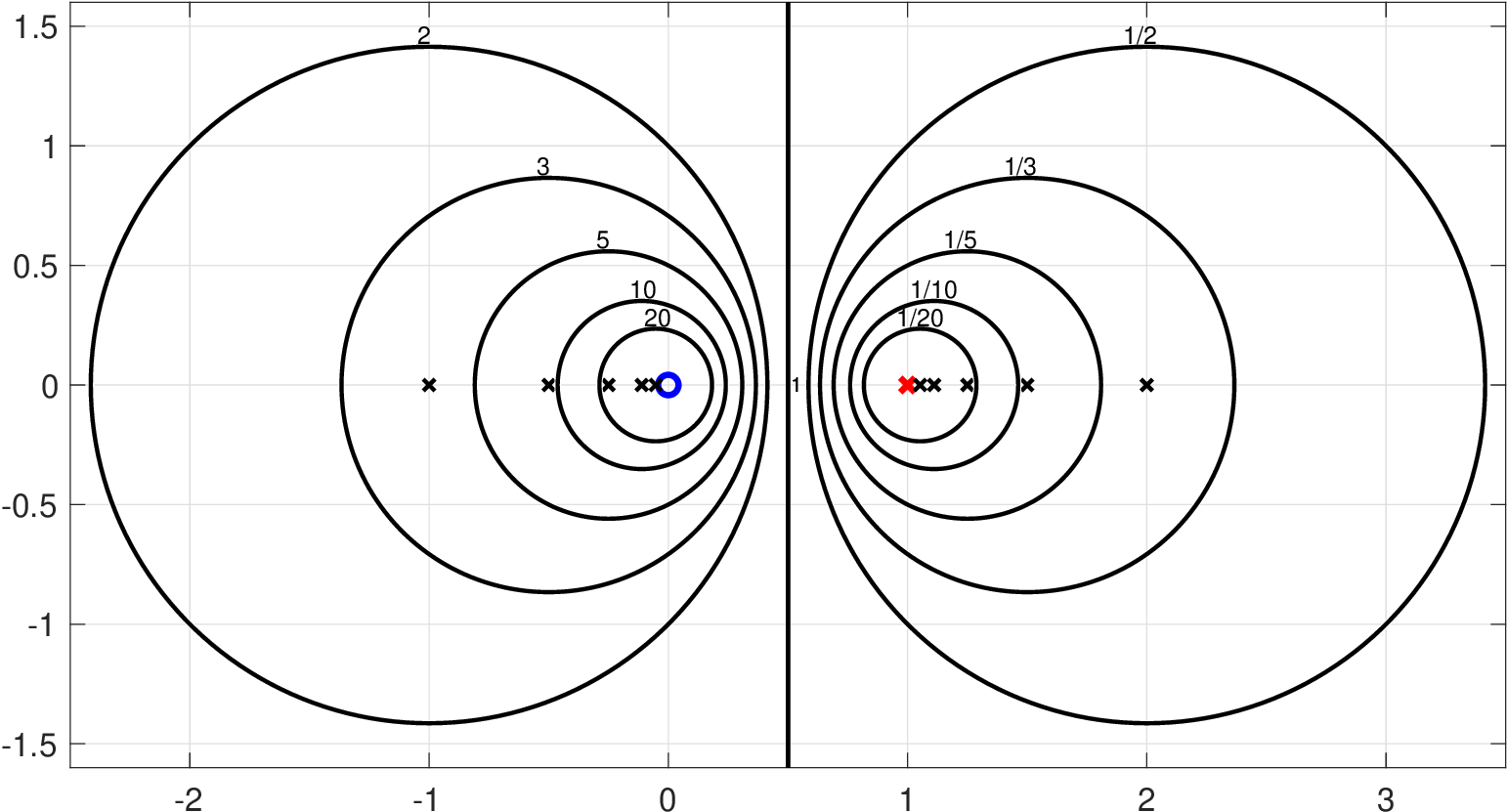,width=\columnwidth}}
\caption{With the desired transmitter at the origin ${\blue \circ}$ and an interferer at ${\red \times}$, the circles represent the
locations where the line-of-sight SIR (squared distance ratio) has values $2,3,5,10,20$. The black crosses are the circle centers. The circles for $\sir$ and $1/\sir$ are mirrored at $x=b/2$.}
\label{fig:illu_single}
\end{figure}

\figref{fig:illu_single} shows the equal distance ratio circles for $b=1$ and $\rho^{2}=2,3,5,10,20$.
Interpreted in a wireless context,
the circles represent the locations on the plane at which an SIR of exactly
$\rho^{2}$ is achieved
in a line-of-sight environment if the desired transmitter is at $(0,0)$ and the interferer at $(b,0)$. 
Letting $r_1=\sqrt{x^2+y^2}$ and $r_2=\sqrt{(x-b)^2+y^2}$ be the distances from $(x,y)$ to $(0,0)$ and $(b,0)$,
respectively, the
SIR is given by the squared distance ratio
\[ \sir_{(x,y)} = \frac{r_1^{-2}}{r_2^{-2}}=\frac{(x-b)^2+y^2}{x^2+y^2} .\]
Hence the locations at which $\sir=\rho^2$ form a circle as given by Lemma \ref{lem:edrc}.

For $\rho>1$, $\sir>\rho^2$ is achieved at locations inside the circle, while for $\rho<1$, it is achieved at locations
outside the circle.
This fact is compactly represented using a M\"obius transform on the complex plane.

\begin{lemma}[M\"obius transform characterization]
Let $\z,\z'\in\C$ denoted the locations of the desired transmitter and interferer in the complex plane.
For $\rho\neq 1$, let 
\begin{equation}
 f_{\rho}(z)\triangleq c+\rho|c|\frac{z-\rho c}{z+\rho c}, \quad z\in\C,
\label{mobi}
\end{equation}
where $c=\|\z'-\z\|/(1-\rho^2)$. 
For $\rho=1$, $f_1(z)\triangleq z+\|\z'-\z\|/2$.

Then, for $\w\in\C$,
\begin{equation}
  \frac{|\w-\z'|}{|\w-\z|}>\rho \quad\Leftrightarrow \quad \w\in \z+f_\rho(H^-)e^{\iu\arg(\z'-\z)}, 
   \label{sirw}
\end{equation}
where $H^-\triangleq \{z\in\C\colon \Re(z)<0\}$ is the open left half-plane of $\C$.
\label{lem:mobi}
\end{lemma}
\begin{IEEEproof}
The modulus of the fraction in \eqref{mobi} is $1$ on the imaginary axis while its argument
ranges from $-\pi$ to $\pi$, so the axis is mapped onto
the circle of radius $\rho|c|$ centered at $c$. The half-plane $H^-$ is mapped inside the circle
if $\rho>1$, since $\Re(z)<0$ decreases the modulus for $c<0$. Conversely,
$\rho<1$, $\Re(z)<0$ increases the modulus since $c>0$.
The rotation and translation in \eqref{sirw} place the disk in the correct position relative to $\z,\z'$.
\end{IEEEproof}
{\em Remarks.}
\begin{itemize}
\item The transform \eqref{mobi} is a generalization of the bilinear transform 
that maps $H^-$ to $|z|<1$
commonly used in digital signal
processing for the continuous time-to-discrete time mapping of
 system functions.
\item Applying the transform to the imaginary axis, with $\z=0$ and $\z'=b$, 
yields the circle in Lemma \ref{lem:edrc} for
$\rho\neq 1$. For $\rho=1$, the definition of $f_1$ ensures that this equivalence extends to the case of
equidistant points to $0$ and $b$, and that $b/(1+\rho)$ lies in the image of the imaginary axis for any $\rho$.
\end{itemize}

\section{Coverage and Stringency}
\subsection{Network Model}
Let $\calX=\{\x_1,\x_2,\ldots\}\subset\R^2$ be a countable set of transmitter locations with at least two elements.
For $\y\in\R^2$, we let $\calX_{-\y}\triangleq\{\x_i-\y\}_{i\in\N}$
be the translated set, and we let $r_i$ be the $i$-th smallest element of $\|\calX_{-\y}\|$,
such that $r_1$ is the distance from $\y$ to the closest transmitter.

With nearest-transmitter association, a path loss exponent $\alpha$, and unit transmit powers, the SIR at location $\y$ is
\[\sir_\y = \frac{h_1 r_1^{-\alpha}}{\sum_{i>1} h_i r_i^{-\alpha}}
 = \frac1{\sum_{i>1} \frac{1}{H_i} \big(\frac{r_1}{r_i}\big)^\alpha}   , \]
where the coefficients $H_i\triangleq h_1/h_i$, $i>1$, are identically distributed with cdf $F_H$.
All $(h_i)_{i\in\N}$ are independent with $\E h_i=1$, and for $i>1$, they are identically distributed.
This formulation allows for different fading distributions for the desired transmitter and the interferers.
In particular, we consider Nakagami fading with parameter $p>0$ for $h_1$ and parameter $q>0$ for $h_i$, $i>1$,
which we refer to as Nakagami-$(p,q)$ fading\footnote{Nakagami-$m$ refers to the symmetric
case $p=q=m$.}.
In this case
\begin{equation}
   F_H(x)=B_{p,q}\left(\frac{px}{px+q}\right) = B_{p,q}\big(\mh{(x\tfrac pq)}\big), 
   \label{gamma_rat}
\end{equation}
where
\[ B_{p,q}(x)\triangleq \frac{\Gamma(p+q)}{\Gamma(p)\Gamma(q)} \int_0^x t^{p-1}(1-t)^{q-1}\dd t, \quad x\in[0,1] ,\]
is the regularized incomplete beta function\footnote{In Matlab, {\tt betainc(x,p,q)}.}, \ie,
the cdf of the beta distribution.
It satisfies $B_{p,q}(1-x)\equiv 1-B_{q,p}(x)$, which reflects the fact that the distribution of $1/H_i$ is related to that
of $H_i$ by flipping $p$ and $q$.
Its inverse\footnote{In Matlab, {\tt betaincinv(u,p,q)}.}
is denoted as $B_{p,q}^{-1}$, \ie, $B_{p,q}(B_{p,q}^{-1}(u))\equiv u$.

{\em Symmetric case: $p=q$.} In the symmetric case, all $(h_i)_{i\in\N}$ are independent and identically distributed (iid),
and  the fading model is the standard Nakagami-$m$ model with $p=q=m$.
For Rayleigh fading $(m=1$), $F_H(x)=\mh{x}$, while for general $m$, the $H_i$ are beta prime distributed with
$F_H(x)=B_{m,m}(\mh{x})$,

{\em Special asymmetric cases:} We consider two special asymmetric cases that can be used to
upper- and lower-bound the performance if the exact
fading distributions are unknown. The first ($p=1$, $q\to\infty$) is pessimistic and assumes Rayleigh fading in the desired
link and no fading in the interfering links. In this case, the $H_i$ are exponential. The second ($p\to\infty$, $q=1$) is
optimistic and assumes no fading in the desired link and Rayleigh fading in the interfering links.
In this case, the $1/H_i$ are exponential. 

As an alternative to the SIR, it is often advantageous to use the {\em signal fraction} (SF)  \cite{net:Haenggi20cl}, defined as the
desired signal power divided by the total received power, \ie, $\sfr=\mh{\sir}$
and 
\begin{equation}
   \{\sfr_\y > t\}\;\; \Leftrightarrow\;\; \{\sir_\y > \imh{t}\},\quad t\in [0,1).
   \label{mh}
\end{equation}
The inverse relationship $t=\mh{\theta}$ can also be expressed as $t=\theta$ MH, where MH
refers to the M\"obius homeomorphic scale that compresses $\R^+$ to $[0,1)$. It offers
advantages for integration and visualization\footnote{See
\url{https://stogblog.net/2020/09/18/unmasking-distributions-with-infinite-support/.} The frequently used dB scale does the opposite---it
extends $(0,1]$ to $(-\infty,0]$, which makes it impossible to draw a complete plot.}
 \cite{net:Haenggi20cl}.

\subsection{Coverage}
\begin{definition}[Coverage]
A location $\y\in\R^2$ is said to be {\em covered} if $\P(\sir_\y>\theta)>u$, where $\theta\geq 0$ and $u\in[0,1)$
are the QoS parameters.
The set of all covered locations is the {\em coverage manifold}
\[ \calC\triangleq \{\y\in\R^2\colon \P(\sir_\y>\theta)>u\} .\]
The set of locations covered when $\x$ is the desired transmitter is denoted by $C_\x$, \ie,
$\calC=\bigcup_{\x\in\calX} C_\x$.
\end{definition}
Frequently a different definition of ``coverage" is employed in the literature, in which $\y$ is ``covered" if
$\sir_\y>\theta$. As pointed out in Subsec.~\ref{sec:approach}, such a definition based on the instantaneous SIR would make coverage vary quickly over time (as the SIR depends on the
small-scale fading), and the covered locations would form a very complex, highly non-contiguous set,
inconsistent with how cellular operators draw coverage maps\footnote{More details on the definitions
of ``coverage"
can be found at \url{https://stogblog.net/2020/09/24/what-is-coverage/}.}.

The coverage manifold critically depends
on a parameter called {\em stringency} that is a function of $\theta$, $u$, and the fading distribution.

\subsection{Stringency}
\begin{definition}[Stringency and QoS regimes]
The {\em stringency} of the QoS pair $(\theta,u)$ is defined as
\begin{equation}
 \sigma(\theta,u) \triangleq \frac{\theta}{F_H^{-1}(1-u)} .\
\end{equation}
If $\sigma<1$, we call the QoS requirements {\em lax}; if $\sigma>1$, they are {\em stringent};
for $\sigma=1$, they are {\em balanced}.
\end{definition}
The stringency is proportional to the SIR threshold parameter $\theta$ and mononotically
increasing with $u$, with $\sigma(\theta,u)\to\infty$ as $u\to 1$.
The next lemma gives concrete expressions for the stringency for different fading models.

\begin{lemma}[Impact of fading on stringency]
For Nakagami-$(p,q)$ fading,
\begin{equation}
  \sigma(\theta,u)=\frac{\theta}{\tfrac qp\imh{(B_{p,q}^{-1}(1-u))}} .
\end{equation}

For symmetric Rayleigh fading ($p=q=1$), this simplifies to 
$\sigma(\theta,u)=\theta\imh{u}$, and for $p=q=1/2$,
\[ \sigma(\theta,u)=\theta\tan^2(\pi u/2). \] 
For increasingly severe Nakagami-$m$ fading, 
\[ \sigma(\theta,u) \overset{m\to 0}{\to}
   \begin{cases} 0 & u\in [0,1/2) \\
                \theta & u=1/2 \\
                \infty & u>1/2 ,
                \end{cases}
\]
while, without fading $(m\to \infty)$, $\sigma(\theta,u)=\theta$.

In the pessimistic asymmetric case ($p=1$, $q\to\infty$),
\[ \sigma(\theta,u)=\frac{\theta}{-\log u} ,\]
while in the optimistic asymmetric case ($p\to\infty$, $q=1$),
\[ \sigma(\theta,u)=-\theta\log(1-u) .\]
\end{lemma}
\begin{IEEEproof}
For Nakagami-$(p,q)$ fading, the quantile function 
\begin{equation}
    F_H^{-1}(u)=\tfrac qp\imh{(B_{p,q}^{-1}(u))}
    \label{nakainv}
\end{equation}
is obtained by inverting \eqref{gamma_rat}. 
For $p=q=1$, this reduces to $F_H^{-1}(u)=\imh{u}$,
and the result follows from $1/\imh{(1-u)}\equiv \imh{u}$.

For $p=q=1/2$, the underlying beta cdf is $B_{1/2,1/2}$, which is the arcsine distribution, 
hence $F_H^{-1}(u)=\imh{\big(\sin^2(\pi u/2)\big)}$, and the result follows from $\tan^2\phi\equiv 1/\imh{(\cos^2(\phi))}$.

For $m\to 0$, we start from the Gauss
hypergeometric function representation \cite[p.~263, (6.6.8)]{net:Abramowitz64} to derive the asymptotic form
\begin{align*}
   B_{m,m}(x)&=\frac{\Gamma(2m)}{\Gamma^2(m)}\frac{x^m}m{}_2F_1(m,1-m;1+m;x) \\
    &\sim \frac m2\frac{x^m}m\left(1+mx+\frac m2 x^2+ \ldots\right),\quad m\to 0\\
    &\sim \frac{x^m}2,\quad m\to 0, \;x\in [0,1).
\end{align*}
Hence $F_H^{-1}(u)\sim \imh{((2u)^{1/m})}$, $u<1/2$, $F_H^{-1}(1/2)=1$ (which holds for all $m>0$), while
$F_H^{-1}(u)\to\infty$ for $u>1/2$,
and it follows that
\[ F_H^{-1}(u) \overset{m\to 0}{\to}
   \begin{cases} 0 & u\in [0,1/2) \\
                1 & u=1/2 \\
                \infty & u>1/2 .
                \end{cases}
\]
Without fading, $H_i\equiv 1$, so $F_H(x)=\one(x\geq 1)$,  $F_H^{-1}(1-u)=\one(u\in [0,1))$, and
$\sigma(\theta,u)=\theta$.

In the asymmetric cases, the results follow from $H_i$ or $1/H_i$ being exponential, respectively.
\end{IEEEproof}

\begin{figure}
\centerline{\epsfig{file=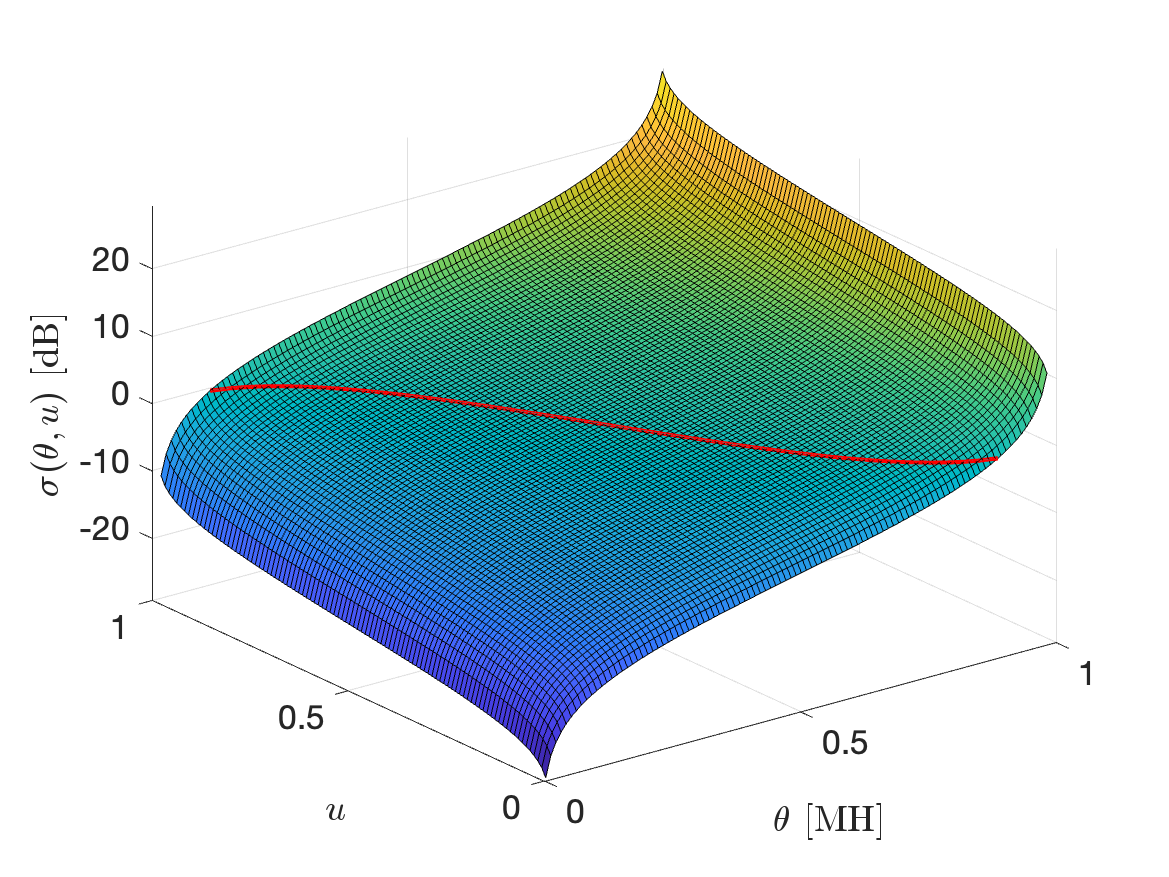,width=\columnwidth}}
\caption{Stringency $\sigma(\theta,u)$ in dB for Nakagami-3 fading. 
To cover the full range $\theta\in\R^+$, the MH scale is used, \ie, $\theta$ MH$=\mh{\theta}\in[0,1)$.
The red line 
is the contour at 0 dB, on which $\theta$ MH $=B_{3,3}^{-1}(1-u)$.}
\label{fig:string}
\end{figure}

\figref{fig:string} shows the stringency $\sigma(\theta,u)$ for Nakagami-$3$ fading, using the MH scale for
$\theta$ to display it in its entire range. 

{\em Remarks.}
\begin{itemize}

\item Comparing the two asymmetric cases, the stringency in the pessimistic case is
significantly larger. The gap between the two diverges as $u\to 1$, and so does the ratio for both $u\to 0$ and $u\to 1$.
For all $u$, the ratio is larger than $2$.

\item In the case of increasingly severe fading, reliabilities less than $1/2$ become trivially satisfied, while
those above $1/2$ become impossible to achieve. The limit as $m\to 0$ can be expected to hold for
general fading models whose variance goes to $\infty$.
\end{itemize}

\section{Bounding the Coverage Manifold Using Q Cells}
\subsection{Q Cells}
\begin{definition}[Q cell]
The Q cell of transmitter $\x\in\calX$, denoted as $Q_\x$, is defined as
\begin{equation}
   Q_{\x}(\rho)\triangleq\Big\{\y\in\R^2\colon \frac{\|\calX\setminus\{\x\}-\y\|}{\|\x-\y\|}>\rho\Big\},\quad \rho>0, 
   \label{defq}
\end{equation}
where $\|\calX\setminus\{\x\}-\y\|$ is the distance to the interferer closest to $\y$.
\end{definition}
The next lemma shows that the Q cell can be analytically described using the M\"obius transform put forth in Lemma \ref{lem:mobi}.
\begin{lemma}[Complex Q cell characterization]
For $\z\in\calX^\sharp$, let
\begin{equation}
   Z_{\z}(\rho) \triangleq \z+\bigcap_{\z'\in\calX^\sharp\setminus\{\z\}} f_\rho(H^-)e^{j\arg(\z'-z)},
   \label{qcell}
\end{equation}
with $f_\rho$ given in \eqref{mobi}. Then $Q_{\x}(\rho)\equiv Z_{\x^\sharp}^\flat(\rho)$.
\end{lemma}
\begin{IEEEproof}
The right side of \eqref{qcell} consists of the locations at which the distance
ratio $|\z'|/|\z|>\rho$ for each interferer $\z'$. This is equivalent to the condition in \eqref{defq} since
if the condition is satisfied for the nearest interferer, it is satisfied for all interferers.
\end{IEEEproof}
{\em Remarks.}
\begin{itemize}
\item For $\rho=1$, the Q cell $Q_\x$ corresponds to the interior of the Voronoi cell $V_\x$,
which is defined as the intersection of the half planes
delineated by the perpendicular bisectors of $\z$ and $\z'$.
\item For $\rho\neq 1$, the Q cell is
the intersection of the regions whose boundaries are the equal distance ratio
circles in Lemma \ref{lem:edrc}.
\begin{itemize}
\item For $\rho>1$, the half planes get morphed into disks per $f_\rho$, and the Q cell is the intersection of these
$|\calX|-1$ disks.
\item For $\rho<1$, the Q cell consists of $\R^2$ (or $\C$) minus a union of exclusion disks for each interferer. 
\end{itemize}
For a given stringency, increasing $\alpha$ reduces $\rho$ if $\rho>1$, resulting in larger Q cell.
If $\rho<1$, increasing $\alpha$ increases $\rho$, which enlarges the exclusions disks.

\item Not every interferer contributes to the shaping of the Q cell. Those that do are termed {\em strong interferers}.

\end{itemize}

\begin{definition}[Strong and dominant interferers]
$\x'$ is a {\em strong interferer} to $\x$ for a given $\rho$ if the removal of $\x'$ increases the Q cell.

A {\em dominant interferer} at location $\y$ is an interferer that by itself is strong enough to cause $\y$ to be uncovered.
\end{definition}

\begin{figure}
\centerline{\epsfig{file=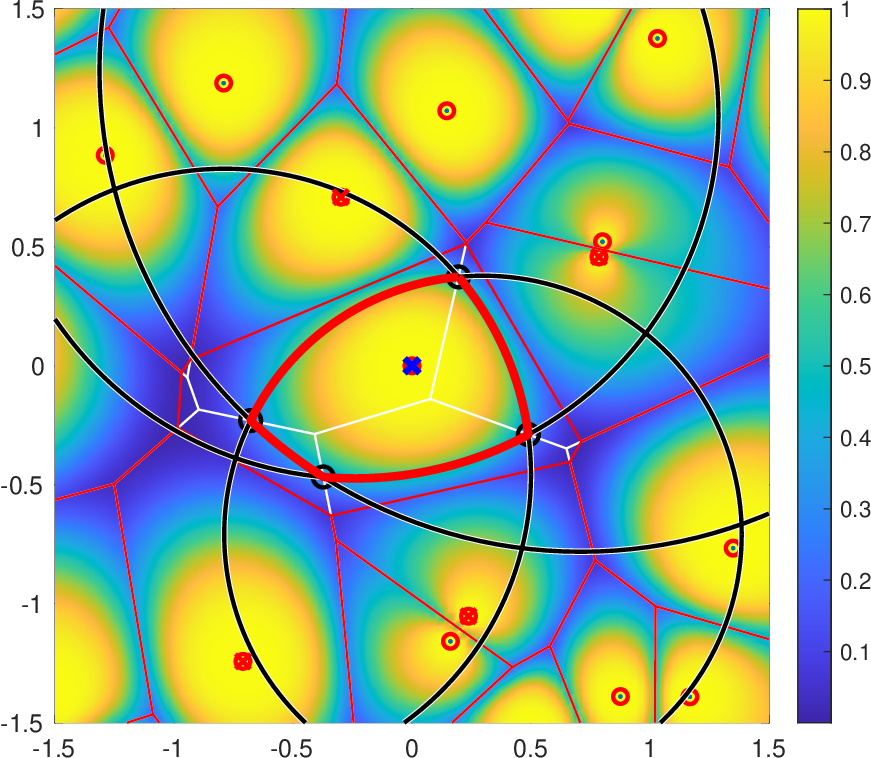,width=.9\columnwidth}}
\caption{Example Q cell for the transmitter at $o$, shown in red. Its boundary is formed by the 
4 black circles, at distance ratio $\rho=\sqrt 2$ with 
respect to each strong interferer whose locations are marked with a red cross.
The red lines are the edges of the Voronoi diagram, and the white lines inside the
central Voronoi cell $V_o$ mark the second-order Voronoi edges.
The entire square region is colored according to $\P(\sir_\y>1)$ for symmetric Rayleigh fading and $\alpha=4$, using the color map on the right. It is apparent that all locations $\y$ for which $\P(\sir_\y>1)>0.8$ are inside the Q disk.}
\label{fig:qcell_map}
\end{figure}

{\em Remarks.}
\begin{itemize}
\item For $\rho<1$, every interferer is strong.
\item For $\rho>1$, the intersection of the $|\calX|-1$ disks in the definition of $Q_\x$ can be reduced to the
intersection of the disks of the strong interferers to $\x$. By Lemma \ref{lem:inter}, their number determines the number of arcs $\na_\x$ that define the boundary of $Q_{\x}$. $\na_\x$ equals the number of Delaunay neighbors of $\x$.
The number of dominant interferers at $\y$
equals the number of disks that $\y$ is outside of. For $\y\in Q_\x$, there are no dominant interferers.
\item For $\rho\downarrow 1$, the number of strong interferers to $\x$ tends to the number of sides
of the Voronoi cell of $\x$. As $\rho\to\infty$, the number may decrease to $1$. 
\item
Denote by $V_{\x\x'}$ the second-order Voronoi cell of $\x$ and $\x'$, \ie, the set of locations that have
$\x$ and $\x'$ as their two closest points in $\calX$ \cite{net:Lee82tcomp,net:Giovanidis15twc}. If 
\[ \y\in (V_{\x}\cap V_{\x\x'})\setminus Q_\x ,\]
$\x'$ is the dominant interferer at $\y$. 
\item By construction, the corner points of the Q cell lie on an edge of the second-order Voronoi diagram.
These are the locations that have exactly two dominant interferers at equal distance. This property will be
exploited in Subsec.~\ref{sec:cutting}. 
\end{itemize}
An example Q cell for $\rho=\sqrt 2$ with the dominant interferers and their equal distance ratio circles is shown in \figref{fig:qcell_map}. As stipulated, the corner points of the Q cell lie on edges of the second-order Voronoi diagram,
which are shown in white.

\begin{figure}
\centerline{\epsfig{file=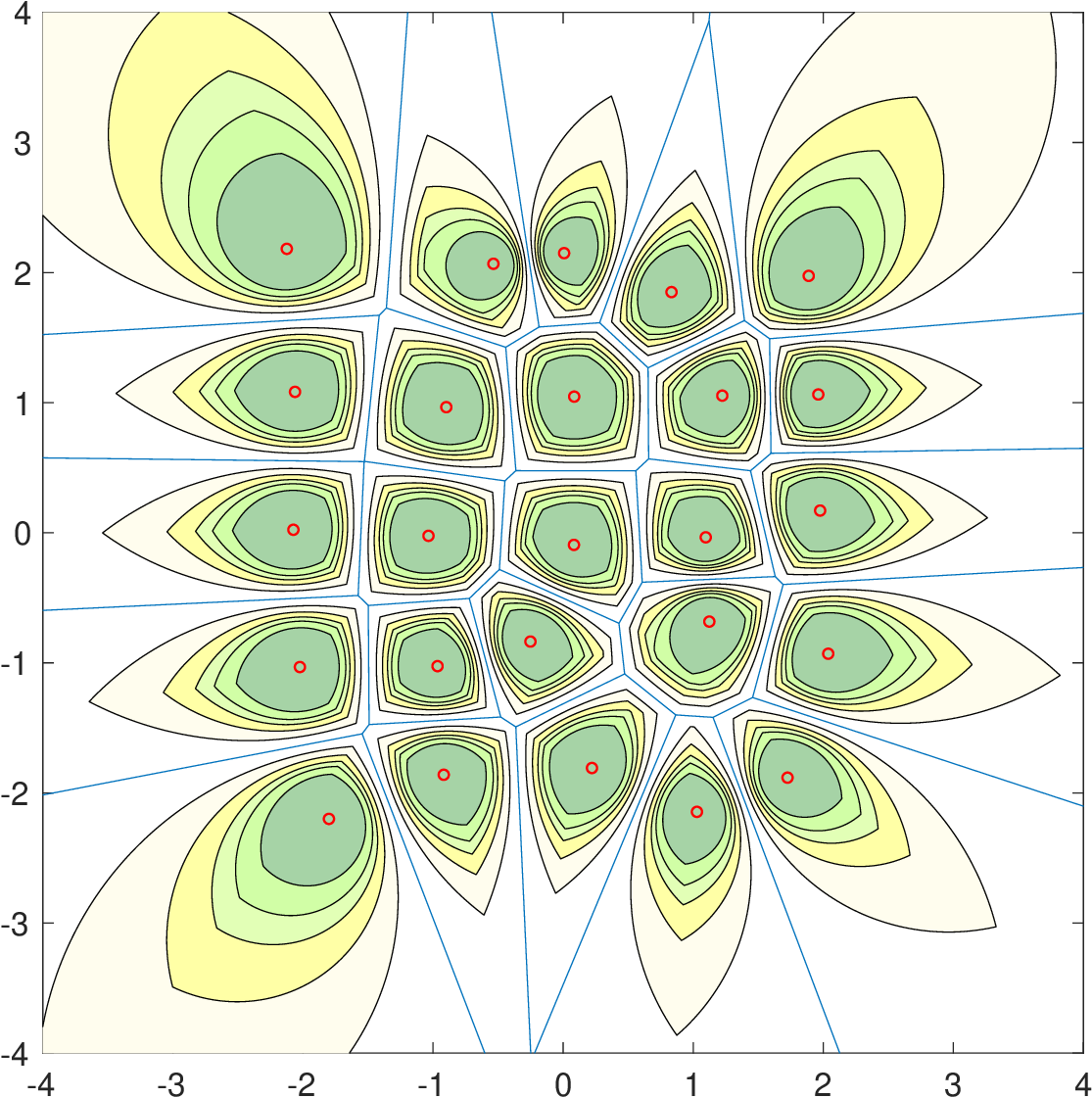,width=.9\columnwidth}}
\caption{Q cells for $\rho=1.25,1.50,1.75,2.00,2.50$ for a set of 25 transmitters. The blue lines are the Voronoi
cell boundaries.}
\label{fig:qcell25}
\end{figure}

\figref{fig:qcell25} shows the Q cells for different values of $\rho$ for a set $\calX$ with 25 elements.

\subsection{Coverage Outer Bounds}
The key missing piece to bound the coverage manifold is the connection between
$\theta$ and $u$ and the radius parameter $\rho$ of the Q cells.
By the following theorem, the connection is established by equating $\rho^\alpha$ with the stringency $\sigma$.
\begin{theorem}[Outer bounds on coverage cells and manifold]
The coverage cell of transmitter $\x$ is bounded as
\begin{equation}
   C_\x\subseteq Q_\x(\rho(\theta,u)), 
   \label{cbound}
\end{equation}
where $\rho(\theta,u)=\sigma(\theta,u)^{1/\alpha}$, 
and the coverage manifold is bounded as
\begin{equation}
   \calC\subseteq \calQ ,
   \label{cmbound}
\end{equation}
where
\[ \calQ\triangleq \bigcup_{\x\in\calX} Q_{\x}(\rho(\theta,u)) \]
is the union of all Q cells.
\label{thm:main}
\end{theorem}
\begin{IEEEproof}
For a transmitter $\x$ and a location $\y$, let $r_1=\|\x-\y|$ and $r_2=\|\calX\setminus\{\x\}-\y\|$ be
the serving distance and nearest interferer distance, respectively. 
With $v\triangleq r_2/r_1$,
\begin{align}
\P\left(\frac{h_1 r_1^{-\alpha}}{h_2r_2^{-\alpha}}>\theta\right) &= \P(Hv^{\alpha} >\theta) \nonumber\\
    &=1-F_H(\theta v^{-\alpha}) .
    \label{reliab}
\end{align}
$1-F_H(\theta v^{-\alpha}) > u$ implies
\begin{align*}
v^{-\alpha} < \frac{F_H^{-1}(1-u)}{\theta}\quad\Longleftrightarrow\quad
 v > \sigma(\theta,u)^{1/\alpha}. 
\end{align*}
Hence a distance ratio smaller than $\rho(\theta,u)$ to the nearest interferer is sufficient
for $\y$ to be uncovered. The bound on the coverage manifold is a consequence of the
bounds of the individual coverage cells.
\end{IEEEproof}
{\em Remarks.}
\begin{itemize}
\item For $\rho>1$, to achieve a reliability $u$, a user needs to be located inside all circles with distance ratio $\rho$
(Lemma \ref{lem:edrc}).

 \item For Rayleigh fading, if coverage is expressed in terms of the SF threshold $t$,
 $\rho(\imh{t},u)^\alpha=\imh{t}\imh{u}$ by \eqref{mh}, which shows a striking symmetry of $\rho$ in the reliability $u$ and the SF threshold $t$.
 The stringent regime is simply $t+u>1$.
 
 \item If only the nearest interferer for each location $\y$ is considered, \eqref{cbound} is an equality.
 \item In the lax regime $\rho< 1$, $\calC\subseteq\calQ$ trivially holds since $\calQ=\R^2$.
For $\rho=1$, $\calQ$ is still essentially $\R^2$ (minus the Voronoi edges).

\item 
Since the Q cells are disjoint in the 
stringent regime, a user cannot be handed over from one transmitter to the next
without losing coverage. Only in the lax regime, seamless coverage is possible.
   
\end{itemize}
The Q cell $Q_o$ in \figref{fig:qcell_map} is drawn for $\rho=\sqrt 2$,
which means it is an outer bound of the coverage cell $C_o$ for all combinations of $\theta$, $u$, and $\alpha$
for which $\sigma(\theta,u)=2^{\alpha/2}$.
For Rayleigh fading and $\alpha=4$, all Q cells are the same if $u=\mh{(4/\theta)}$.
The color map of the figure indicates $\P(\sir_\y>1)$ for each $\y$ in the square region, and the
corresponding reliability is $u=\mh{4}=0.8$.
 Hence outside the Q cell, in $V_o\setminus Q_o$, it is guaranteed that $\P(\sir_\y>1)<0.8$, which can be
 confirmed from the colors---none of these locations reach a shade of yellow.

\subsection{Cutting Corners}
\label{sec:cutting}
Here we focus on the stringent regime $\rho>1$ and show that is is possible to improve on the outer coverage bounds
by ``cutting the corners" of the Q cells. The reason why the corner points of the Q cells are of special
interest is that they are equidistant to the two nearest interferers, while the definition of the Q cell
only considers a single nearest interferer. By taking into account the joint interference from the
two equidistant interferers, we can obtain a better outer bound of the coverage provided by each transmitter,
\ie, a {\em refined Q cell}.

First we define a version of a Q cell that always considers the two nearest interferers.

\begin{definition}[Second-Order Q Cell]
For $|\calX|>2$,
the second-order Q cell $Q_{\x}^{(2)}$ is the set of location that are covered if the two nearest interferers to each location
are considered. Formally, 
\begin{align}
   Q_{\x}^{(2)} &\triangleq \Big\{\y\in\R^2\colon \P\left(\frac{h_1r_1^{-\alpha}}{h_2 r_2^{-\alpha}+h_3 r_3^{-\alpha}}>\theta\right)>u\Big\} \nonumber \\
     &=\Big\{\y\in\R^2\colon \P\left(\frac1{H_2 v_2^\alpha}+\frac1{H_3 v_3^\alpha}<\frac1{\theta}\right)> u\Big\},
     \label{q2_def}
\end{align}
where $(r_i)_{i\in [3]}$ are the ordered distances to $\y$ and $v_i=r_i/r_1$ and $H_i=h_1/h_i$,  as before.
\label{def:second}
\end{definition}
{\em Remarks.}
\begin{itemize}
\item By construction, $C_\x\subseteq Q^{(2)}_\x\subseteq Q_\x$.
\item  This definition could be extended easily to account for more than $2$ interferers by adding more terms
$1/(H_i\rho_i)$ to \eqref{q2_def}, but the higher-order Q cells
are increasingly intractable. 
\end{itemize}
The key purpose of introducing the second-order cell is the definition of a refined version of the Q cell if $\na_\x>1$, \ie,
if the Q cell has corner points. Before doing so, we need to introduce another variant of the Q cell.

\begin{definition}[Interference-cloned Q Cell]
Let
\begin{equation}
   \rho^*(\theta,u) \triangleq \left(\frac{\theta}{F_{H^*}^{-1}(1-u)}\right)^{1/\alpha}, 
   \label{rhostar}
\end{equation}
where $F^{-1}_{H^*}$ is the quantile function of $H^*\triangleq h_1/(h_2+h_3)=H_2H_3/(H_2+H_3)$.
The Q cell $Q_\x(\rho^\ast)$ is the {\em interference-cloned} version of the standard Q cell $Q_{\x}(\rho)$.
\end{definition}
The interference-cloned Q cell is the Q cell obtained if every interferer is replaced by two interferers at the same distance, each with independent fading or, equivalently, if every interferer has two antennas, sufficiently spaced
for the fading to be independent and each transmitting at unit power. The ratio
\[ \frac{\rho^*}{\rho} = \left(\frac{F_H^{-1}(1-u)}{F_{H^*}^{-1}(1-u)}\right)^{1/\alpha} \]
is larger than one since $F_{H^*}(x)>F_H(x)$ and thus $F^{-1}_{H}(1-u)>F^{-1}_{H^*}(1-u)$.
For Nakagami-$(p,q$) fading,
 \begin{equation}
   F_{H^*}(x)=B_{p,2q}\left(\frac{px}{px+q}\right)=B_{p,2q}\big(\mh{(x\tfrac pq)}\big), 
   \label{gamma_rat2}
\end{equation}
which is a slight modification of \eqref{gamma_rat}.
For $p=q=1$, $F_{H^*}(x)=x(x+2)/(x+1)^2$ and thus $F^{-1}_{H^*}(u)=(1-u)^{-1/2}-1$.
It follows that $(\rho^*)^\alpha=\theta\imh{\sqrt u}$ and
$\rho^*/\rho=\left(\frac{\imh{\sqrt u}}{\imh{u}}\right)^{1/\alpha}$.

The next lemma establishes a connection between $Q^{(2)}(\rho)$ and $Q(\rho^*)$.

\begin{lemma}[Corner points of the interference-cloned cell]
The corner points of $Q_{\x}(\rho^*)$ (if any) lie on the boundary $\partial Q^{(2)}_\x$ of the second-order cell.
\end{lemma}
\begin{IEEEproof}
For the interference-cloned cell,
\begin{align}
\P\left(\frac{h_1 r_1^{-\alpha}}{(h_2+h_3)r_2^{-\alpha}}>\theta\right) &= \P(H^* v^\alpha >\theta) 
\label{condic}\\
    &=1-F_{H^*}(\theta v^{-\alpha}) \nonumber,
\end{align}
as in the proof of Thm.~\ref{thm:main} but with $H$ replaced by $H^\ast$.
Hence a distance ratio of $(\theta/F^{-1}_{H^*}(1-u))^{1/\alpha}$ demarcates the boundary of the interference-cloned cell $Q_\x(\rho^*)$.

Now, at the corner points $Q_\x(\rho^*)$, there are two actual equidistant interferers (without cloning), so 
at these locations, cloning has the same effect as taking the two nearest interferers in account, hence
they lie on $\partial Q^{(2)}(\rho)$ and have $\P(\sir_\y>\theta)\leq u$, with equality if there are only two
interferers.
\end{IEEEproof}

The remaining question is how to ``interpolate" between $Q_\x(\rho)$ and $Q_\x(\rho^*)$ to define a refined Q cell.

\begin{definition}[Refined Q Cell]
Let $\rho^*(\theta,u)$ be given as in \eqref{rhostar}
and denote by $\q_k$, $k\in[\na_\x]$, the corner points of $Q_{\x}(\rho^*)$
and by $\phi_k$ and $\phi'_k$ the angles of the tangents to the two arcs meeting at $\q_k$.
Letting $\bar\phi_k=(\phi_k+\phi'_k)/2$, we define
\[ \calH_\x\triangleq \bigcap_{k=1}^{\na_\x} \q_k^\sharp+H^- e^{\iu(\bar\phi_k-\pi/2)} \]
to be the intersection of the half-planes whose boundary is the line through $\q_k$ at angle $\bar\phi_k-\pi/2$,
such that $\x^\sharp\in\calH_\x$. If $\na_\x=0$, $\calH=\C$, per the usual convention that the intersection
of $0$ sets is the universal set.

The refined Q cell is
\[ Q^*_\x(\rho) \triangleq Q_{\x}(\rho)\cap \calH_\x^\flat. \]
\end{definition}

\begin{figure}
\centerline{\epsfig{file=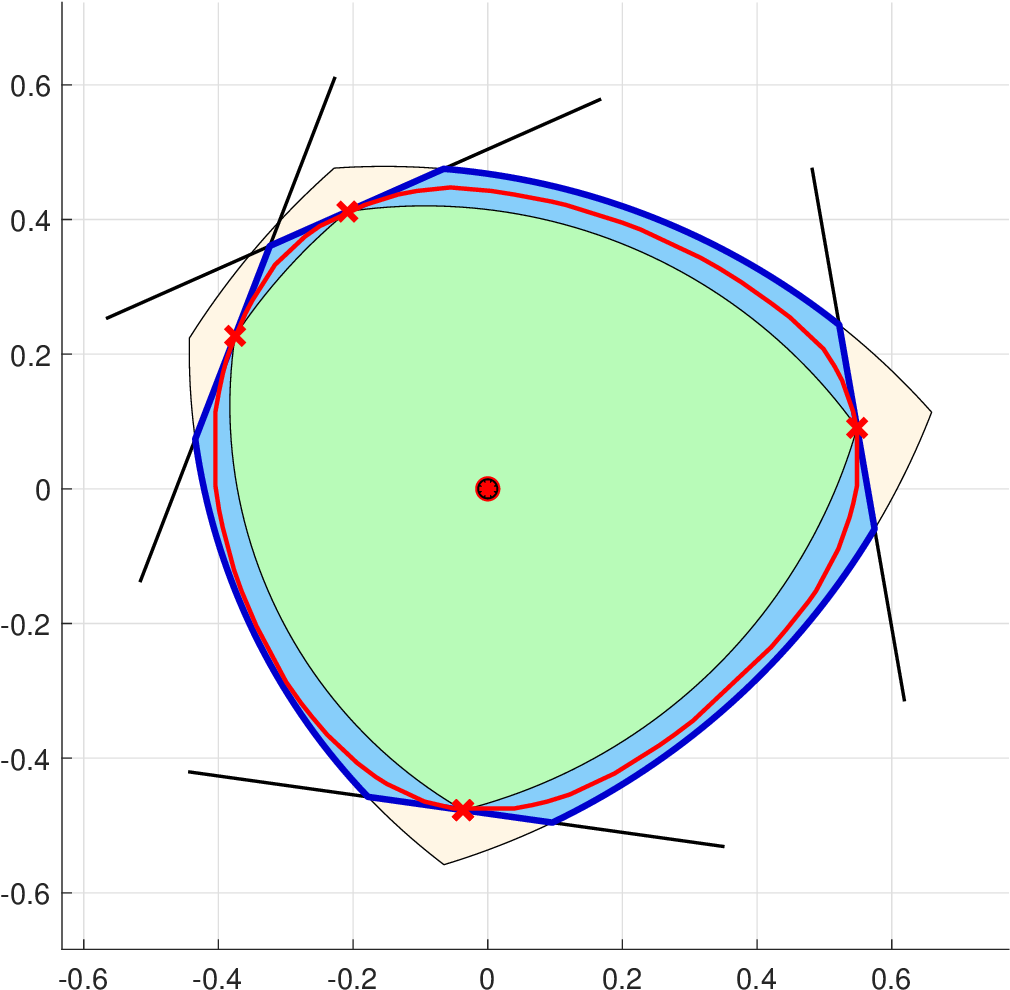,width=.8\columnwidth}}
\caption{Example Q cell $Q(\rho)$ with $\rho=\sqrt 3$ (outermost cell, including the beige parts)
and corresponding interference-cloned cell $Q(\rho^*)$, where
$\rho^*=2.07$, (innermost cell, green) and refined Q cell $Q^*(\rho)$ (blue boundary),
and the simulated boundary of the second-order cell $Q^{(2)}$ (red line). The
corners of $Q(\rho^*)$, marked with red crosses, lie on the boundary of $Q^{(2)}$.
The black line segments mark the boundaries of the half-planes touching the corners of $Q(\rho^*)$, effectively
``cutting the corners" off of $Q(\rho)$. In this example, the refined Q cell is 8\% smaller than the original Q cell.}
\label{fig:qstar_example}
\end{figure}

\begin{figure}
\centerline{\epsfig{file=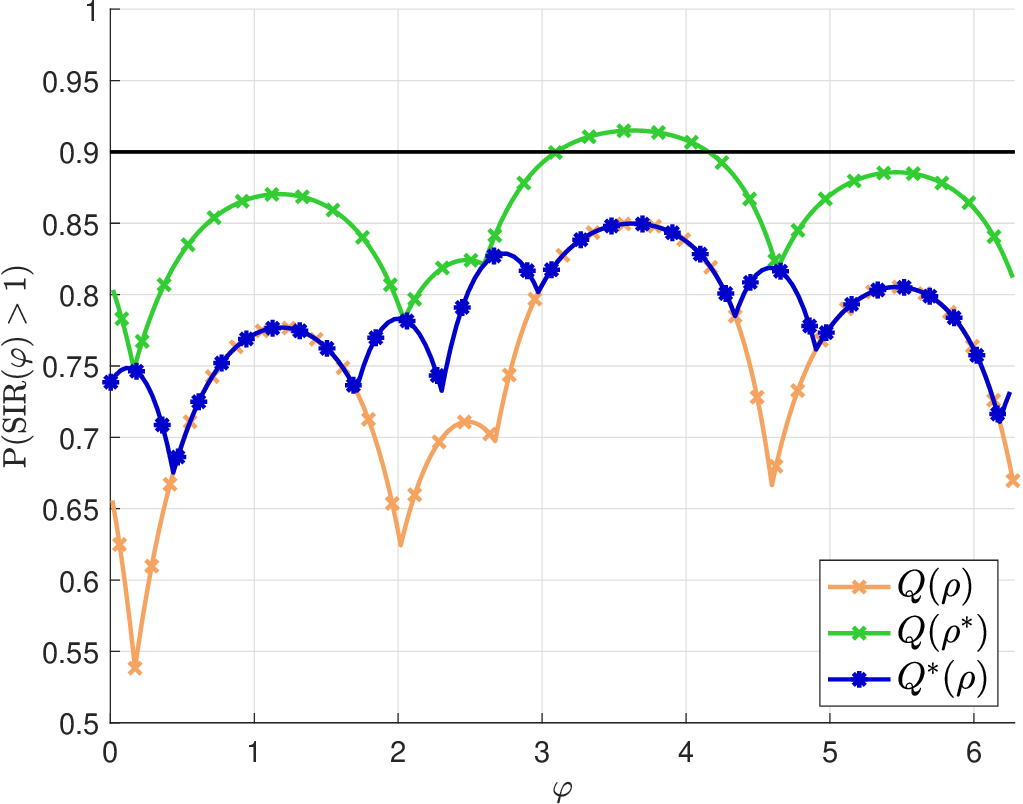,width=.85\columnwidth}}
\caption{Reliability on the boundary of the Q cells in \figref{fig:qstar_example}, with the colors
matching those of the corresponding cells in \figref{fig:qstar_example}. The boundary is represented in polar
coordinates relative to the transmitter $\x$. The black line indicates the target reliability $u=0.9$,
corresponding to $\rho=(\imh{u})^{1/4}=\sqrt 3$ for Rayleigh fading, $\theta=1$, and $\alpha=4$.}
\label{fig:qcell_ps}
\end{figure}
As before we set $\calQ^*\triangleq\bigcup_{\x\in\calX} Q^*_\x$.

\figref{fig:qstar_example} shows an example of a Q cell $Q(\rho)$, its interference-cloned version $Q(\rho^*)$,
and the refined Q cell $Q^*(\rho)$, together with the simulated boundary of the $Q^{(2)}$ cell. It is apparent that the corners of $Q^*(\rho)$ lie on the boundary of $Q^{(2)}$, hence the corners of $Q(\rho)$ can be cut up to those inner
corners.

\begin{theorem}[Improved outer bound]
The refined Q cell $Q^*_\x$ is a tighter outer bound on $C_\x$ than $Q_\x$,
\ie,
\begin{equation}
   C_\x \subseteq Q^*_\x \subseteq Q_\x \quad \forall \x\in\calX,
   \label{ref_bound}
\end{equation}
and $\calC\subseteq\calQ^*\subseteq\calQ$.
\label{thm:main2}
\end{theorem}
\begin{IEEEproof}
First inclusion: Since $Q^{(2)}_\x$ is convex, the intersection with a tangential half-plane at any point of its boundary
leaves it unchanged. So $Q^{(2)}_\x\subseteq Q^*_\x$, which implies $C_\x\subseteq Q^*_\x$.
Second inclusion: Since the corner points of $Q^*_\x$ lie inside $Q_\x$, the intersection with $\calH_\x^\flat$
makes $Q^*_\x$ smaller than $Q_\x$---unless there are no corner points.
\end{IEEEproof}
If there are more than two interferers, the first inclusion in \eqref{ref_bound} becomes strict, and
if $Q_\x(\rho^*)$ has at least two corner points, the second inclusion becomes strict.

\figref{fig:qcell_ps} shows the probabilities $\P(\sir>1)$ on the boundary of the Q cells in \figref{fig:qstar_example}.
The boundary is represented in polar coordinates relative to the transmitter $o$. The Q cells are constructed
for a target probability $u=0.9$. It is apparent that the refined Q cell has a significantly higher reliability at the
angles of the corner points of the standard Q cell.

\section{Coverage Estimation in Infinite Networks}
In this section, we focus on the stringent QoS regime and
assume that $\calX$ is countably infinite, in such as way that it forms a {\em proper point pattern}
as per \cite[Def.~1]{net:Haenggi24wcl}. This means its convex hull spans $\R^2$ and all Voronoi cells
are finite. If $\calX$ is random, \ie, a point process, these properties are assumed to hold in an almost sure sense.

\subsection{Covered Area Fraction}
A key quantity in infinite networks is the covered area fraction.
\begin{definition}[Area fraction covered by a set]
For $B\subset\R^2$, 
\[ \eta_B\triangleq \lim_{r\to\infty}\frac{|B\cap b(o,r)|}{\pi r^2} \in [0,1]\]
is the area fraction covered by $B$.
\end{definition}
An immediate consequence of Theorems~\ref{thm:main} and \ref{thm:main2} is $\eta_\calQ\geq \eta_{\calQ^*}\geq \eta_\calC$.
In the next two subsections, we quantify $\eta_{\calQ}$ for different types of deployments $\calX$.

For the asymptotic scaling of $\eta_\calQ$, we have the following lemma. We define $\delta\triangleq 2/\alpha\in (0,1)$.
\begin{lemma}[Asymptotic coverage of \boldmath$\eta_\calQ$]
For all stationary and ergodic point processes, $\eta_\calQ(\rho)=\Theta(\rho^{-2})=\Theta(\theta^{-\delta})$, $\rho\to\infty$.
\label{lem:asymp}
\end{lemma}
\begin{IEEEproof}
As $\rho\to\infty$, all equal distance ratio circles of $\x\in\calX$ become concentric at $\x$, and their radii are proportional
to the distance from $\x$ to the interfering transmitters and decreasing with $\rho^{-1}$. Hence asymptotically
only transmitters with minimum distance, given by $\argmin_{\x'\in\calX\setminus\{\x\}} \|\x-\x'\|$ characterize the
Q cell. If there is a single interferer at the minimum distance, the Q cell is a disk with area proportional to
$\rho^2/(\rho^2-1)^2=\Theta(\rho^{-2})$ by  Lemma \ref{lem:edrc}. If there are multiple interferers at the minimum
distance, \ie, if $\calX$ forms a lattice, the Q cell is the intersection of a fixed number of
disks with equal radius, whose area also
scales with $\Theta(\rho^{-2})$. This behavior applies to all Q cells, hence the covered area fraction scales in the
same way.
Lastly, since $\rho^\alpha\propto\theta$ by Thm.~\ref{thm:main}, $\rho^{-2}\propto \theta^{-\delta}$.
\end{IEEEproof}

\subsection{Lattices}
In regular lattices, all Voronoi cells $V_\x$ are congruent, and so are all Q cells $Q_{\x}$.
Hence their area fractions can be calculated by considering just a single cell.
\begin{lemma}[Q cell area fraction in lattices]
For the square lattice,
\begin{equation}
   \eta_{\calQ^{\square}}(\rho)= \frac{4\rho^2\arctan\left(\frac{\sqrt{2\rho^2-1}-1}{\sqrt{2\rho^2-1}+1}\right)-2\sqrt{2\rho^2-1}+2}
{(\rho^2-1)^2} .
\label{sq_lattice}
\end{equation}
For the triangular lattice,
\begin{equation}
   \eta_{\calQ^{\hexagon}}(\rho)= 
\frac{4\sqrt 3\rho^2\arctan\left(\frac{r-\sqrt 3}{\sqrt 3 r+1}\right)-\sqrt 3 r+3}{(\rho^2-1)^2},
\label{tri_lattice}
\end{equation}
where $r=\sqrt{4\rho^2-1}$.
For both square and triangular lattices,
\begin{equation}
\eta_\calQ(\rho) < \frac4{(1+\rho)^2} .
\label{up_lattice}
\end{equation}
\label{lem:lattices}
\end{lemma}
\begin{IEEEproof}
\eqref{sq_lattice} and \eqref{tri_lattice} are obtained from straightforward calculation of the areas of the disk segments and the
inner square and hexagon, respectively.
For the square lattice, the tightest square outer bounding the Q cell
has side length $2/(1+\rho)$; for the triangular lattice,
the tightest hexagon outer bounding the Q cell has inner radius $a=\sqrt 2/(3^{1/4}(1+\rho))$, hence its area $6a^2/\sqrt 3=4/(1+\rho)^2$ is the same.
\end{IEEEproof}
Since the arctan in \eqref{sq_lattice} tends to $\pi/4$, $\eta_{\calQ^{\square}}\sim \pi \rho^{-2}$, $\rho\to\infty$.
For the triangular lattice, the arctan tends to $\pi/6$, so $\eta_{\calQ^{\hexagon}}\sim (2\pi/\sqrt 3) \rho^{-2}$, which is $15\%$ larger. The upper bound \eqref{up_lattice} is asymptotically larger by a factor $2\sqrt{3}/\pi\approx 1.10$ than the Q cells for the triangular lattice, so the bound is at most $10\%$ off.

\subsection{Point Processes}
If $\calX$ is a stationary and ergodic point process of density $\lambda$, 
\[ \eta_\calQ=\lambda \E_o (|Q_o(\rho)|) ,\]
where $\E_o$ is the expectation with respect to the Palm measure of $\calX$ at $o$ \cite[Ch.~8]{net:Haenggi12book}.
In words, the area fraction of $\calQ$ is the mean area of the typical Q cell divided by the mean area $1/\lambda$
of the typical Voronoi cell.

\begin{figure}
\centerline{\epsfig{file=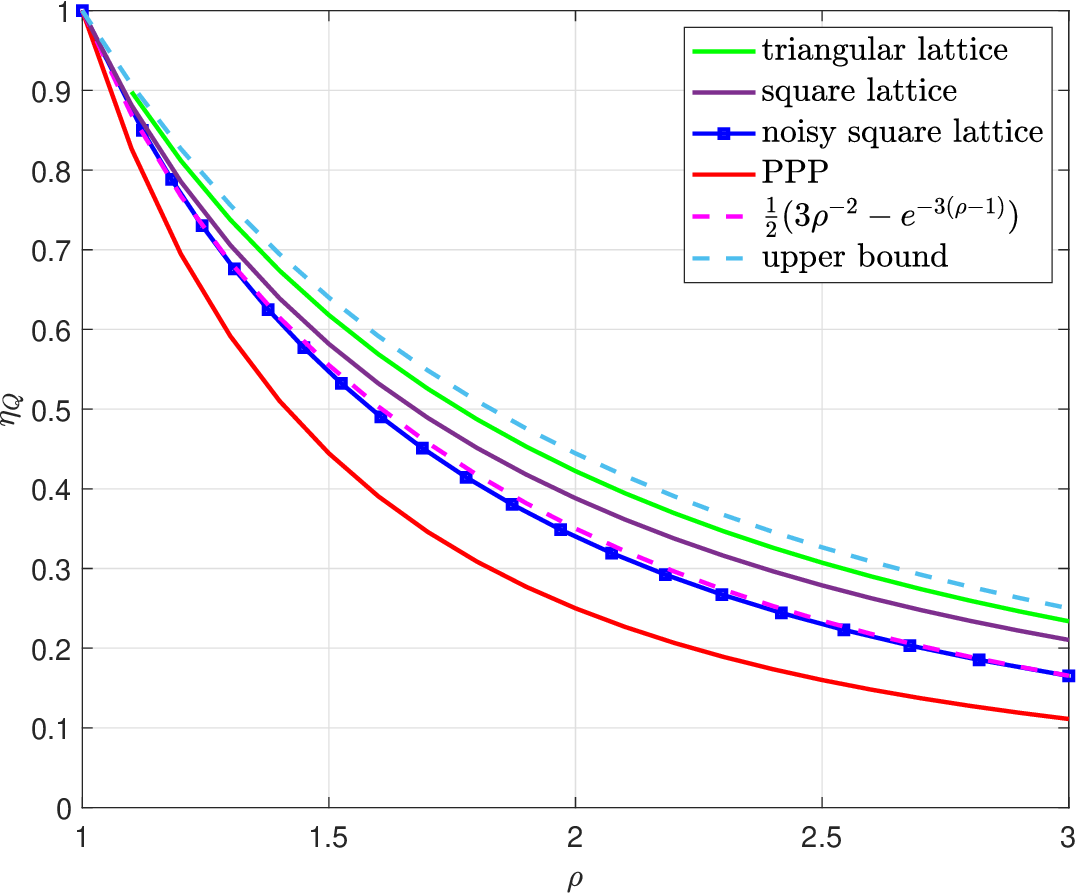,width=.9\columnwidth}}
\caption{Area fraction $\eta_\calQ$ for the triangular lattice \eqref{tri_lattice}, the square lattice \eqref{sq_lattice}, the PPP $\rho^{-2}$ (from Lemma \ref{lem:cov_ppp}), and a noisy square lattice with variance $1/16$ (simulated). The upper bound is from \eqref{up_lattice}, and the approximation 
\eqref{eta_approx} is also shown.}
\label{fig:area_frac}
\end{figure}
The following lemma gives the area fraction of $\calQ$ for the Poisson point process (PPP).
\begin{lemma}[Q cell area fraction for PPP]
For the PPP, $\eta_\calQ=\rho^{-2}$.
\label{lem:cov_ppp}
\end{lemma}
\begin{IEEEproof}
The area fraction only depends on $\rho$, so the result follows from re-writing the MD for the nearest interferer case
in  \cite[Cor.~3]{net:Haenggi21cl2} in terms of $\rho$.
\end{IEEEproof}
This lemma generalizes \cite[Eqn.~(8)]{net:Haenggi21cl2} to arbitrary fading, including asymmetric cases,
and it establishes which combinations of $\theta$, $u$, and $\alpha$ result in the same Q cell coverage.

\figref{fig:area_frac} shows the area fractions for the square lattice, the noisy square lattice where a Gaussian
perturbation with variance $1/16$ is added independently to both coordinates of each lattice point, and the PPP.
\begin{table}
\[ \begin{array}{|c|c|c|}
\hline
\text{point process} & \rho\downarrow 1 & \rho\to\infty \\\hline
\text{triangular lattice} & 1-\frac{10}9(\rho-1) & \frac{2\pi}{\sqrt 3}\rho^{-2} \\
\text{square lattice} & 1-\frac43(\rho-1) & \pi \rho^{-2} \\
\text{PPP} & 1-2(\rho-1) & \rho^{-2} \\\hline
\end{array} \]
\caption{Asymptotics of $\eta_\calQ(\rho)$ for three types of deployments from Lemmas \ref{lem:lattices} and
\ref{lem:cov_ppp}.}
\label{table:asym}
\end{table}
Table \ref{table:asym} shows the asymptotic behavior of $\eta_\calQ(\rho)$ for three point processes.

Since practical deployments often are about in the middle of lattices and PPPs in terms of regularity,
a good general approximation is
$\eta_\calQ(\rho)\approx f_\calQ(\rho)$ for
\begin{equation}
f_\calQ(\rho)= \frac32\rho^{-2}-\frac12e^{-3(\rho-1)}.
\label{eta_approx}
\end{equation}
This function has slope $-3/2$ at $\rho=1$ and decays as $f_\calQ(\rho)\sim \frac32\rho^{-2}$ as $\rho\to\infty$.
It is apparent from \figref{fig:area_frac} that it matches the area fraction of the noisy lattice quite exactly.

\begin{theorem}[Uncovered users]
For any independent stationary and (jointly) ergodic point processes of transmitters and users, 
 the fraction $\mu$ of uncovered users is lower bounded by
\begin{equation}
\mu > 1-\frac4{(1+\rho)^2} .
\label{uncovered}
\end{equation}
If the transmitters form a PPP, the fraction of uncovered users is larger than $1-\rho^{-2}$.
\label{thm:uncov}
\end{theorem}
\begin{IEEEproof}
The triangular lattice is the optimum deployment, for which the Q cell area is upper bounded by \eqref{up_lattice}.
The result for the PPP follows from Lemma \ref{lem:cov_ppp}.
\end{IEEEproof}
The theorem establishes that for $\rho=4/\sqrt 3-1\approx 1.31$, at least $25\%$ of the users are uncovered, and for
$\rho=2\sqrt 2-1\approx 1.83$, at least $50\%$ of the users are uncovered.
For the PPP, the two thresholds are $\rho=2/\sqrt 3\approx 1.155$ and $\rho=\sqrt 2$, respectively.

\subsection{Connection to the SIR Meta Distribution}
We first state an important fact about the SIR meta distribution (MD) 
and the covered area fraction $\eta_\calC$.
\begin{fact}[Meta distribution and \boldmath$\eta_\calC$]
If $\calX$ is a stationary and ergodic point process, 
\[ \eta_\calC(\rho(\theta,u))=\md(\theta,u) ,\]
where $\md(\theta,u)\triangleq\P(\P(\sir_o>\theta\mid\calX)>u)$ is the meta distribution of the
SIR \cite{net:Haenggi16twc,net:Haenggi21cl1}.
\label{fact:md}
\end{fact}
Hence, the SIR MD can be either interpreted as $\eta_\calC$ or
as the fraction of users who achieve an SIR of $\theta$
with probability at least $u$
for a stationary and ergodic point process of users.

Consequently, $\eta_\calQ$ and $\eta_{\calQ^*}$ are upper bounds on the meta distribution.
In particular, \eqref{uncovered} is a universal upper bound for all point processes and all values of
$\theta$ and $u$, as stated in the next corollary.
\begin{corollary}[Upper bounds on MD]
For any independent stationary and ergodic point processes of transmitters and users, the MD is bounded as
\begin{equation}
   \md(\theta,u) < \frac4{(1+\rho)^2} =\frac4{\left(1+\Big(\frac{\theta}{F_H^{-1}(1-u)}\Big)^{1/\alpha}\right)^2} . 
   \label{md_bound}
\end{equation}
For the PPP,
\begin{equation}
  \md(\theta,u)<(F_H^{-1}(1-u))^\delta \theta^{-\delta} .
  \label{md_bound_ppp}
\end{equation}
\label{cor:md_bound}
\end{corollary}
\begin{IEEEproof}
Follows from Thm.~\ref{thm:uncov} and Fact \ref{fact:md}.
\end{IEEEproof}
{\em Remarks.}
\begin{itemize}
\item For Rayleigh fading, the upper bound \eqref{md_bound_ppp} simplifies to $(\theta\imh{u})^{-\delta}$.
\item If the MD exceeds the right side of \eqref{md_bound_ppp}, the transmitters form a more regular point process than the PPP.
\item If only the nearest interferer is considered in the MD and coverage definition,
$\md(\theta,u)=\eta_\calC=\eta_\calQ(\rho(\theta,u))$. Lemma \ref{lem:cov_ppp} is a consequence
of this equality---the result follows from the nearest interferer-only version of the MD.
\end{itemize}
\subsection{Separability in Poisson Networks}
If $\calX$ is a stationary PPP,
the stringent regime corresponds to the {\em separable regime}
where, for iid fading, $\md(\theta,u)=g(u)\theta^{-\delta}$ \cite[Thm.~2]{net:Feng20twc}. ``Separable"
refers to the fact that the MD can be written as a product of a function of $u$
and $\theta^{-\delta}$.
The bound \eqref{md_bound_ppp} establishes that $g(u)<(F_H^{-1}(1-u))^\delta$.
For Nakagami-$m$ fading,
the stringent/separable regime is
 $\{(\theta,u)\colon u>1-B_{m,m}(\mh{\theta})\}$.

For Rayleigh fading, $\md(\theta,u)\sim \sinc\delta(\theta\imh{u})^{-\delta}$, $u\to 1$, \cite[Cor.~5]{net:Feng20twc},
which suggests that 
\begin{equation}
  \eta_\calC\approx \sinc\delta\, \eta_\calQ.
  \label{approx_scaling}
\end{equation}  

The asymptotic proportionality of the MD to $\theta^{-\delta}$ as $\theta\to\infty$
holds for all stationary point processes with Rayleigh fading \cite[Thm.~3]{net:Feng20twc}.
For $\eta_\calQ$, it holds for all types of fading.

Only for the PPP,  $\eta_\calQ$ is separable. For all point processes,
{\em asymptotic separability} holds in the sense that 
$\eta_\calQ(\rho)\sim\tilde g(u)\theta^{-\delta}$, $\theta\to\infty$, where
$\tilde g(u)=\lim_{\theta\to\infty}\eta_\calQ(\rho)\theta^\delta$,
which follows from Lemma \ref{lem:asymp}. It is expected that asymptotic separability
also holds for the MD.

\subsection{Scaled Q Cells for Coverage Manifold Estimation}
Q cells only account for one interferer, while refined Q cells account for two in specific locations.
Shrinking the Q cells can approximately incorporate the effect of the remaining interferers.
If the MD (\ie, $\eta_\calC$) is available, the Q cells can be shrunk to match the area fraction.
Applied to the refined cells, 
letting $\nu=\eta_\calC/\eta_{\calQ^*}$, each cell is scaled as
\[ \hat Q^*_\x\triangleq \x+\sqrt\nu(Q^*_\x-\x), \]
and we let $\hat\calQ^*\triangleq\bigcup_{\x\in\calX} \hat Q^*_\x$.
This way, $\eta_{\hat\calQ^*}=\eta_\calC$.

To assess the accuracy of $Q^*_\x$ and $\hat Q^*_\x$ as an outer bound and approximation of $C_\x$, 
respectively, we use the
probabilities $\P(\sir>\theta)$ averaged over the Q cell boundaries. Formally,
\begin{align}
  U_\x&\triangleq \frac1{|\partial Q^*_\x|} \int_{\y\in\partial Q^*_\x} \P(\sir_\y>\theta)\dd\y, \\
  \hat U_\x&\triangleq \frac1{|\partial \hat Q^*_\x|} \int_{\y\in\partial \hat Q^*_\x} \P(\sir_\y>\theta)\dd\y .
\end{align}
For random $\calX$, the ccdfs of $\sir_\y$ are conditioned on $\calX$, and $U$ and $\hat U$ are random
variables. Without subscript, $U$ and $\hat U$ refer to the average probabilities on the boundary of the typical
cells $Q^*_o$ and $\hat Q^*_o$, respectively.

Along the boundary $\partial C_\x$, the reliability is $u$ by definition, and,
since $C_\x\subset Q^*_\x$, $U_\x<u$. Hence the proximity of $U_\x$ to $u$ and the deviation
of $\hat U_\x$ from $u$ are sensible indicators for the accuracy of the Q cells.

The effect of the scaling can be viewed as an increase in $\rho$ and, in turn,
an increase of the reliabilities (or the threshold $\theta$) on the cell boundaries.
Using $\eta_{\calQ^*}\approx \rho^{-2}$, which holds in general, we have $\hat \rho^{-2}=\nu\rho^{-2}\approx\eta_\calC$,
so $Q^*_\x(\hat\rho)\approx \hat Q^*_\x(\rho)$ for $\hat\rho=\rho/\sqrt{\nu}$.
Then, from \eqref{reliab}, 
\begin{equation}
   \hat u\triangleq 1-F_H(\theta\hat\rho^{-\alpha})=1-F_H(\theta\nu^{1/\delta}\rho^{-\alpha})
   \label{uhat}
\end{equation}
is an estimate of the maximum reliability on the boundary of $\hat Q^*_\x$.
For Rayleigh fading, $\hat u=1-\mh{(\theta\nu^{1/\delta}\rho^{-\alpha})}=\mh{(\nu^{-1/\delta}\imh{u})}$,
where $u$ is the target reliability of the unscaled Q cells.

\subsubsection{Square lattice}
\begin{figure}
\centerline{\epsfig{file=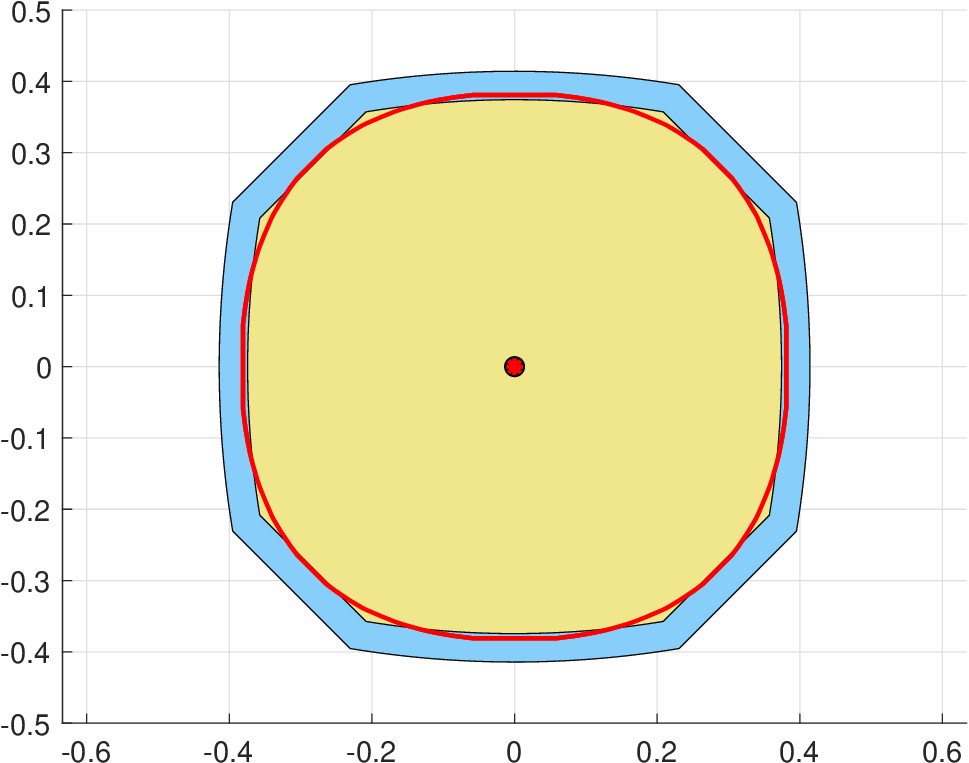,width=.7\columnwidth}}
\caption{Refined Q cell (outer cell, including blue part) and scaled refined Q cell (inner yellow cell) and boundary
of coverage cell (red line) for $\rho=\sqrt 2$ for the square lattice. The area scaling factor from the refined
Q cell $Q^*_o$ to the actual cell $C_o$ is $\nu=0.817$.}
\label{fig:scaled_qstar_square}
\end{figure}

\begin{figure}
\centerline{\epsfig{file=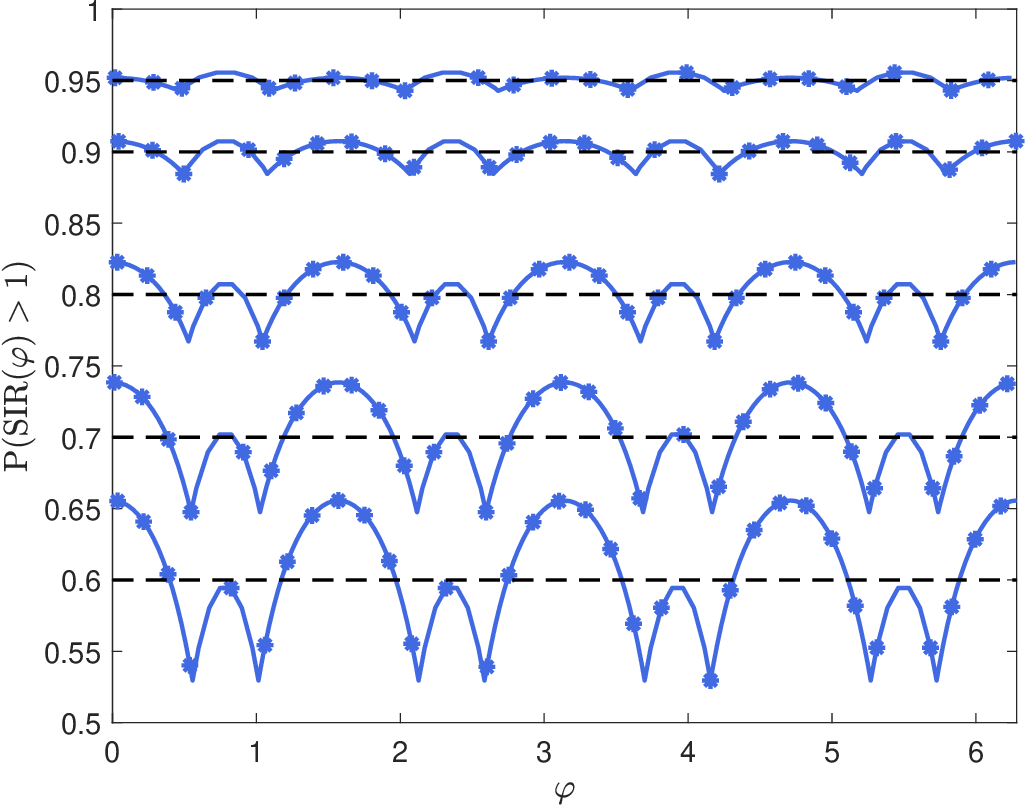,width=.9\columnwidth}}
\caption{Probability $\P(\sir>1)$ along the boundary of the scaled refined cell $\hat Q^*$ in the square lattice
for $\rho=\sigma(1,u)^{1/4}=(\imh{u})^{1/4}$, $u\in\{0.6,0.7,0.9,0.95\}$, with Rayleigh fading. The values of $u$ are indicated using dashed lines.}
\label{fig:ps_scaled_qstar}
\end{figure}

\figref{fig:scaled_qstar_square} displays the refined Q cell $Q^*_o$, the scaled version $\hat Q^*_o$, and the
true coverage cell $C_o$ for the square lattice for $\rho=\sqrt 2$.
\figref{fig:ps_scaled_qstar} shows the probabilities $\P(\sir>1)$ along the boundary $\partial\hat Q^*_o$
for targets $u\in\{0.6,0.7,0.8,0.9,0.95\}$.
It is apparent that with increasing $u$, the probabilities are increasingly well concentrated around $u$,
from $0.6\pm 0.05$ to $0.95\pm 0.005$. The averages over the boundary are
$0.606, \: 0.703, \: 0.801, \: 0.949$, which are very close to the target values of $u$.

\subsubsection{Poisson point process}
\begin{figure}
\centerline{\epsfig{file=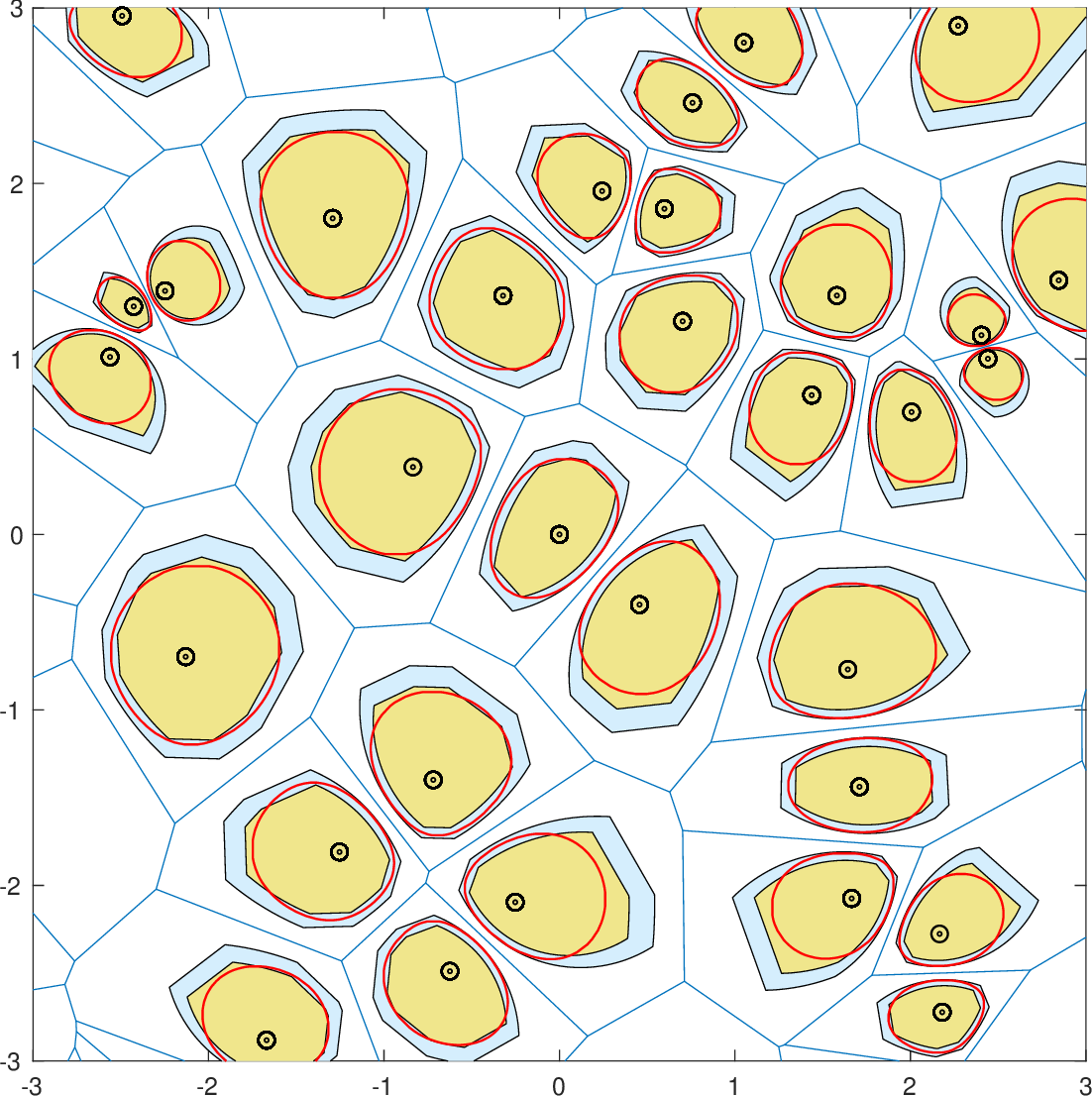,width=\columnwidth}}
\caption{Refined Q cells (outer cells, including blue part) and scaled refined Q cells (inner yellow cells) 
for $\rho=\sqrt 2$ and boundaries of
the coverage cells $C$ (red lines) for $\theta=1$, $u=0.8$, $\delta=1/2$, and Rayleigh
fading for a realization of a PPP. The area scaling factor from the refined to the actual cell is $\nu=0.676$.}
\label{fig:scaled_qstar_ppp}
\end{figure}

\figref{fig:scaled_qstar_ppp} shows refined Q cells $Q^*_\x$, scaled refined Q cells 
$\hat Q^*_\x$, and coverage cells $C_\x$ for a realization of a PPP of intensity $1$ for Rayleigh fading.

From $\hat Q^*_\x\approx C_\x$ follows that $\E(\hat U)\approx u$, and,
from \eqref{uhat}, $\E(\hat U)\approx \mh{(\nu^{-1/\delta}\imh{(\E(U))})}$,
 which can be inverted to yield
\begin{equation}
  \E(U)\approx \mh{(\nu^{1/\delta}\imh{u})}.
  \label{rel_approx}
\end{equation}  

\begin{figure}
\centerline{\epsfig{file=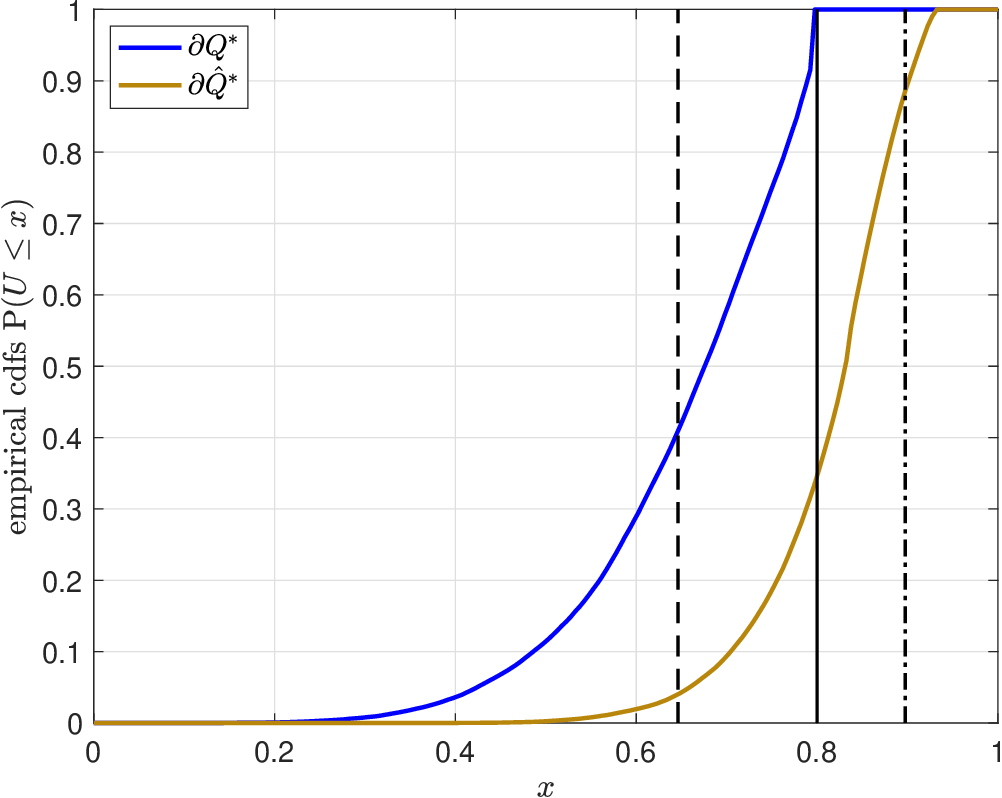,width=.9\columnwidth}}
\caption{Empirical cdfs of $\P(\sir>1\mid\calX)$ for the PPP on the boundaries $\partial Q^*_o$ and $\partial\hat Q^*_o$.
The vertical lines indicate the target $x=u=0.8$ (solid line) and $\mh{(\nu^{1/\delta}\imh{u})}=0.646$ (dashed line). The
dashed-dotted line at $x=0.898$ is the estimate \eqref{uhat} of the upper bound of the reliabilities.
$\E(\hat U)=0.821$ and $\E(U)=0.654$.} 
\label{fig:ppp_cdf}
\end{figure}

\figref{fig:ppp_cdf} shows the empirical cdfs $\P(U\geq x)$ and $\P(\hat U\geq x)$ for $\theta=1$, estimated from
1000 Q cells with target reliability $u=0.8$ ($\rho=\sqrt 2$) in a realization of a PPP. The means are $\E(\hat U)=0.821$, which is close to $u$ as stipulated, and $\E(U)=0.654$, which is close to
$\mh{(\nu^{1/\delta}\imh{u})}=0.646$, in agreement with \eqref{rel_approx}.
The empirical variances are 
$\var(U)\approx 0.0149$ and $\var(\hat U)\approx 0.0069$, so the values on $\partial\hat Q^*$
are well concentrated in the interval $[\mh{(\nu^{1/\delta}\imh{u})},\mh{(\nu^{-1/\delta}\imh{u})}]$,
which is the interval between the dashed and dash-dotted lines in \figref{fig:ppp_cdf}.

\subsubsection{Area scaling without MD}
If the MD is not available, the scaling factor $\nu$ can be set 
heuristically as 
\begin{equation}
   \nu(\delta,g) = (\sinc\delta)^{1-g(1-\delta)} ,
   \label{nu_heuristic}
\end{equation}
where $g$ quantifies the regularity of the point process. For a lattice, $g=1$, while for the PPP, $g=0.3$.
For perturbed lattices with variance $v$, $g=\max\{0.3,1-3.5 v\}$.

The rationale behind \eqref{nu_heuristic} is the following:
\begin{itemize}
\item Since both $\eta_\calC$ and $\eta_{\calQ^*}$ scale similarly with $\rho$, $\rho$ does not need to be included as an explicit
parameter in an estimate of $\nu$.
\item If $\delta\to 0$, $Q^*_\x\to C_\x$ for all $\x$, \ie, $\nu(0)=1$. If $\delta\to 1$, the interference becomes
overwhelming, and $|C_\x|\to 0$, \ie,
$\nu(1)=0$. For intermediate $\delta$, $\nu(\delta)$ is monotonically decreasing.
\item The base $\sinc\delta$ is taken from \eqref{approx_scaling}, which is raised to an exponent
that depends on the regularity $g$ and $\delta$ and takes into account that \eqref{approx_scaling} is
the approximate scaling factor of the standard Q cells, not of the refined ones.
For higher regularity, the exponent is smaller since
the refined Q cell is more accurate.
\item For $\delta\ll 1$ ($Q^*\approx C$ in all cases) and $1-\delta\ll 1$ ($|C|\approx 0$ in all cases),
the regularity of the point process has little influence. To reflect this, $g$ is multiplied by $1-\delta$.
\end{itemize}

\begin{figure}
\centerline{\epsfig{file=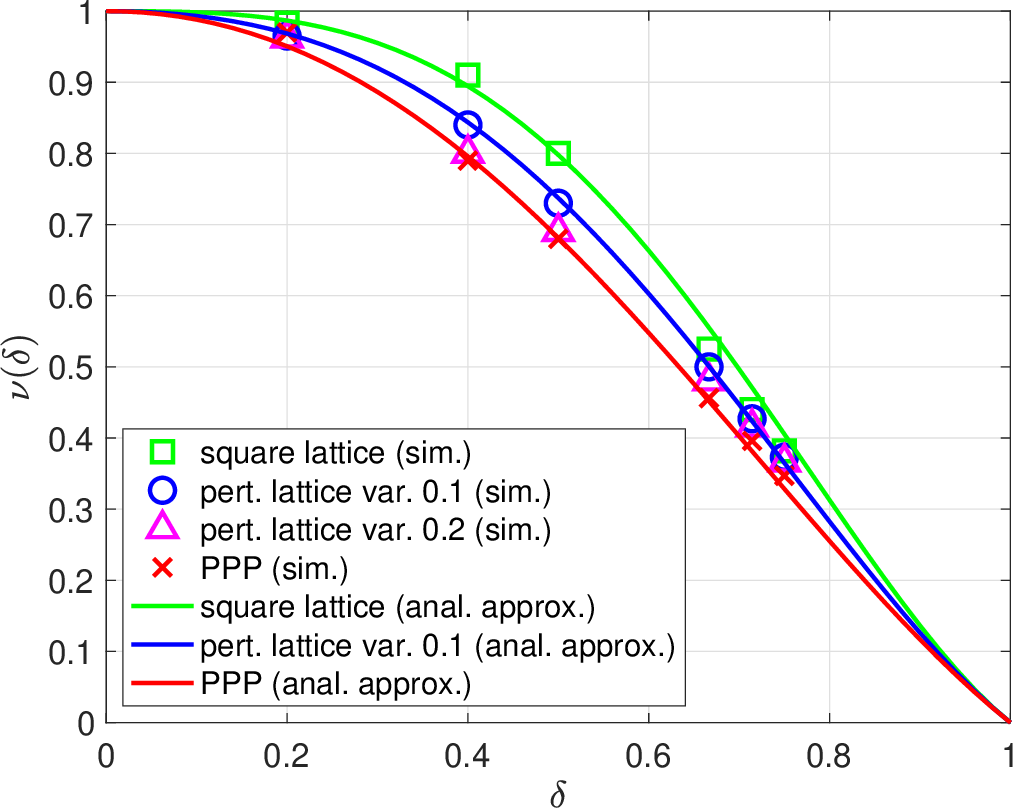,width=.85\columnwidth}}
\caption{Area scaling factor $\nu(\delta)=\eta_\calC/\eta_{\calQ^*}$ for different deployments. The markers
are obtained from simulations, the solid lines from \eqref{nu_heuristic}.}
\label{fig:area_corr}
\end{figure}

\figref{fig:area_corr} compares this simple heuristic with simulation results for the square lattice,
perturbed lattices, and the PPP.

In the example of $\delta=1/2$ used in \figref{fig:scaled_qstar_ppp},
$(\sinc(1/2))^{0.85}\approx 0.68$, which is close to the true scaling of $0.676$ (see the red line and
the red $\times$ in \figref{fig:area_corr} at $\delta=1/2$).

We summarize the steps to obtain the scaled refined Q cells in the following recipe.
\begin{recipe}[Scaled refined Q cells]
\leavevmode
\begin{itemize}
\item[(1)] For given $\theta $ and $u$, calculate the stringency $\sigma(\theta,u)$ and set $\rho=\sigma^{1/\alpha}$.
\item[(2)] Determine the refined Q cells $Q^*_\x(\rho)$ for each $\x\in\calX$.
\item[(3a)] If the MD $\eta_\calC$ is available, set $\nu=\eta_\calC/\eta_{\calQ^*}$ and scale each $Q^*_\x$ to $\hat Q^*_\x$
so that $\eta_{\hat\calQ^*}=\eta_\calC$.
\item[(3b)] If the MD is not available, set $\nu$ according to \eqref{nu_heuristic} and 
scale each $Q^*_\x$ to $\hat Q^*_\x$ so that
$\eta_{\hat\calQ^*}\approx \eta_\calC$.
\end{itemize}
\label{recipe}
\end{recipe}

\section{Conclusions}
{\em Q cell analysis (QCA)} is a powerful method to bound and approximate the coverage manifold in
wireless networks.
Standard Q cells constitute simple yet fairly accurate outer bounds on the coverage region of each transmitter.
They depend on the QoS parameters $\theta$ and $u$ only through the scalar stringency $\sigma$ and can be
calculated very efficiently as the intersection of a modest number of disks whose boundaries are formed
by equal distance ratio circles with ratio $\rho=\sigma^{1/\alpha}$.
As a generalization of Voronoi cells, they are defined as the intersection of M\"obius transformed half-planes
instead of just the intersection of half-planes. In the stringent regime, the complement of their union is
guaranteed to be uncovered. Cutting corner points to
account for equidistant interferers results in refined Q cells that tightly outer bound the coverage regions.

The stringency has a simple analytical expression for general Nakagami-$(p,q)$ fading, which allows
for different fading types in the serving and interfering links.
It establishes that for all combinations of $\theta$, $u$, and $\alpha$ that
have the same distance ratio $\rho=\sigma^{1/\alpha}$, the coverage performance will be similar.
Accordingly, it is unnecessary to explore many combinations of the three parameters---the coverage outer
bound only depends on the single scalar parameter $\rho$.

Applied to infinite networks,
Q cell coverage is much more tractable than the actual coverage (or SIR meta distribution) but closely related.
It allows for the derivation of simple universal bounds. For instance,
in any infinite network with $\rho=2\sqrt 2-1$, at least 50\% of the users are uncovered.
In Poisson networks, more than 50\% are uncovered at $\rho=\sqrt 2$.

We expect that QCA will be useful to quickly explore the effect of advanced
transmission schemes such as coordinated multipoint (CoMP) or non-orthogonal multiple access (NOMA),
and the effects of resource allocation, scheduling and load balancing---and to facilitate their implementation.
For instance, users outside the Q cells are natural candidates for bandwidth aggregation, BS silencing or CoMP.

\bibliographystyle{IEEEtr}

 \end{document}